\documentclass[
 reprint,onecolumn,
 amsmath,amssymb,
 aps,
]{revtex4-1}

\usepackage{epsfig,bm,epsf,graphics}
\usepackage{graphicx,subfigure}
\usepackage{dcolumn}
\usepackage{bm}
\usepackage{xcolor}

\begin{document}
\preprint{APS/123-QED}

\title{PHASE TRANSITION IN FRUSTRATED THIN FILMS - PHYSICS AT PHASE BOUNDARIES}
\author{H. T. Diep}
\affiliation{%
Laboratoire de Physique Th\'eorique et Mod\'elisation,
Universit\'e de Cergy-Pontoise, CNRS, UMR 8089\\
2, Avenue Adolphe Chauvin, 95302 Cergy-Pontoise Cedex, France.\\
 }%

\date{\today}
\begin{abstract}
In this review, I outline some principal theoretical knowledge on the properties of frustrated systems and thin films.
The two points I would like to emphasize: i) the physics in low dimensions where exact solutions can be obtained, ii) the physics at phase boundaries where spectacular phenomena can occur due to competing interactions of the two phases around the boundary.  This competition causes a frustration.
I will concentrate my attention to thin films and phenomena occurring near  the boundary of two phases of different symmetries.
The case of two-dimensional (2D) systems is in fact the limiting case of thin films with a monolayer. Naturally, I will treat this case at the beginning.
After a short introduction on frustrated spin systems, I show several 2D frustrated Ising
spin systems which can be exactly solved by using vertex models. These systems contain most of the spectacular
effects due to the frustration: high ground-state degeneracy, existence of several phases in the ground-state phase
diagram, multiple phase transitions with increasing temperature, reentrance, disorder lines, partial disorder at
equilibrium. Evidences of such effects in non solvable models are also shown and discussed.
Thin films are next presented with different aspects: surface elementary excitations (surface spin-waves), surface phase transition and criticality.
Several examples are shown and discussed.  New results on skyrmions in thin films and superlattices are also displayed.
\vspace{0.5cm}
\begin{itemize}
\item PACS numbers: 75.10.-b ; 75.10.Hk ; 64.60.Cn
\item Keywords: Frustration; Phase Transition; Reentrance ; Disorder Lines;\\
 \hspace {2cm}  Surface Spin-Waves ; Thin Films; Theory; Simulation.
\end{itemize}
\end{abstract}

\maketitle

\section{Introduction}
Materials science has been extensively developed during the last 30 years. This is due to an enormous number of industrial applications which drastically change our life style.  The progress in experimental techniques, the advance on theoretical understanding and the development of high-precision simulation methods together with the rapid increase of computer power have made  possible the spectacular development in materials science. Today, it is difficult to predict what will be discovered in this research area in ten years.

The purpose of this review is to look back at early and recent results in the physics of frustrated systems at low dimensions: 2D systems and thin films.
We would like to connect these results, published over a large period of time, on a line of thoughts: physics at phase boundaries.  A boundary between two phases of different orderings is determined as a compromise of competing interactions each of which favors one kind of ordering.  The frustration is thus minimum on the boundary (see reviews on many aspects of frustrated systems in Ref. \cite{DiepFSS}).  When an external parameter varies, this boundary changes and we will see in this review that many interesting phenomena occur in the boundary region.   We will concentrate ourselves in the search for interesting physics near the phase boundaries in various frustrated systems in this review.

The study of order-disorder phase transition is a fundamental task of equilibrium statistical mechanics \cite{DiepSP,Zinn}. Great efforts have been
made to understand the basic mechanisms responsible for spontaneous ordering as well as the nature of the phase transition
in many kinds of systems.   We will show methods to study the phase transition in thin films where surface effects when combined with frustration effects give rise to many new phenomena.  Surface physics has been intensively developed in the last two decades due to its many applications \cite{zangwill,bland-heinrich,DiepTM,Baibich,Grunberg,Fert,Fert2013}.

A large part of this review, section \ref{2DI}, is devoted to the definition of the frustration and to models which are exactly solved.  We begin with exactly solved models in order to have all properties defined without approximation. As seen, many spectacular phenomena are exactly uncovered such as partial disorder, reentrance, disorder lines and multiple phase transitions.  Only exact mathematical techniques can allow us to reveal such beautiful phenomena which occur around the boundary separating two phase of different ground-state orderings. These exact results permit to understand similar behaviors in systems that cannot be solved such as 3D systems.

In section \ref{thinfilms}, an introduction on surface effects in thin films is given. In order to avoid a dispersion of techniques, I introduce only the Green's function method  which can be generalized in more complicated cases such as non-collinear spin states.  Calculations of the  spin-wave spectrum and the surface magnetization are in particular explained.

In section \ref{surfaces} a number of striking results obtained mainly by my group are shown on a number of frustrated thin films including helimagnetic films.   We show in particular the surface phase transition, quantum fluctuations at low temperature, and the existence of partial phase transition. Results obtained by Monte Carlo simulations are also shown in most cases to compare with the Green's function technique.

The question of  the criticality in thin films is considered in section \ref{critical}. Here, the high-precision multi-histogram techniques are used to show that critical exponents in thin films are effective exponents having values between those of the 2D and 3D universality classes.

Section \ref{skyrmions} is devoted to skyrmions, a hot subject at the time being due to their numerous possible applications. Here again, we show only results obtained in the author's group, but we mention a large bibliography.  Skyrmions are topological excitations. They are a kind of circular domain walls involving a number of spins.   Skyrmions are shown to result from the competition of different antagonist interactions under an applied magnetic field. We  find the existence of a skyrmion crystal, namely a network of periodically arranged skyrmions. Results show that such a skyrmion crystal is stable up to a finite temperature.  The relaxation time of skyrmions is shown to follow a stretched exponential law.

Concluding remarks are given in section \ref{Concl}.

\section{Physics in two dimensions: frustration effects}\label{2DI}
\subsection{Frustration}
During the last 30 years, much attention has been paid to frustrated models \cite{DiepFSS}. The concept of "frustration" has been introduced \cite {Tou,Villain1} to describe the situation where a spin (or a number
of spins) in the system cannot find an orientation to {\it fully} satisfy all the interactions with its neighboring spins
(see below). This definition can be applied to Ising spins, Potts models and vector spins. In general, the frustration is
caused either by competing interactions (such as the Villain model \cite {Villain1}) or by lattice structure as in the
triangular, face-centered cubic (fcc) and hexagonal-close-packed (hcp) lattices, with antiferromagnetic nearest-neighbor
(nn) interaction. The effects of frustration are rich and often unexpected (see \cite{DiepFSS}).

In addition to the fact that real magnetic materials are often frustrated due to several kinds of interactions, frustrated spin systems have their own interest in statistical mechanics.
Recent studies show that many established statistical methods and theories have encountered many difficulties in dealing
with frustrated systems. In some sense, frustrated systems are excellent candidates to test approximations and improve
theories.

Since the mechanisms of many phenomena are not understood in real systems (disordered systems, systems with
long-range interaction, three-dimensional systems, etc), it is worth to search for the origins of those phenomena in
exactly solved systems.  These exact results will help to understand qualitatively the behavior of real systems which are
in general much more complicated.

Let us  give here some basic definitions to help readers unfamiliar with these subjects.

\begin{figure}[h!]
\centering
\includegraphics[width=2 in]{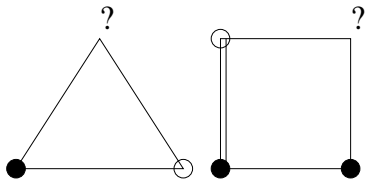}
\caption{Examples of frustrated  plaquettes:  ferro- and antiferromagnetic
interactions, $J$ and $-J$, are shown by single and double lines, $\uparrow$ and $\downarrow$ Ising spins by black and void
circles, respectively. Choosing any orientation for the spin marked by the question mark will leave one of its bonds
unsatisfied (frustrated bond).} \label{fig:IntroFP}
\end{figure}

Consider two spins $\mathbf S_i$ and $\mathbf S_j$ with an interaction $J$. The interaction energy is $E=-J \left(\mathbf
S_i \cdot \mathbf S_j\right)$. If $J$ is positive (ferromagnetic interaction) then the minimum of $E$ is $-J$ corresponding
to the configuration in which $\mathbf S_i$ is parallel to $\mathbf S_j$. If $J$ is negative (antiferromagnetic
interaction), the minimum of $E$ corresponds to the configuration where $\mathbf S_i$ is antiparallel to $\mathbf S_j$.  It
is easy to see that in a spin system with nn ferromagnetic interaction, the ground state (GS) of the system corresponds to
the spin configuration where all spins are parallel: the interaction of every pair  of spins is fully satisfied. This is
true for any lattice structure. If $J$ is antiferromagnetic, the spin configuration of the GS depends on the lattice
structure: i) for lattices containing no elementary triangles, i.e. bipartite lattices (such as square lattice, simple
cubic lattices, ...) the GS is the configuration in which each spin is antiparallel to its neighbors, i.e. every bond is
fully satisfied. ii) for lattices containing elementary triangles such as the triangular lattice, the fcc lattice and the
hcp lattice, one cannot construct a GS where all bonds are fully satisfied (see Fig. \ref{fig:IntroFP}). The GS does not
correspond to the minimum of the interaction of every spin pair. In this case, one says that the system is frustrated.

The first frustrated system which was studied in 1950 is the triangular lattice with Ising spins interacting with each
other via a nn antiferromagnetic interaction \cite{Wan}. For vector spins, non collinear spin configurations due to
competing interactions were first discovered in 1959 independently by Yoshimori \cite{Yos}, Villain \cite{Vill} and
Kaplan \cite{Kapl}.

Consider an elementary cell of the lattice. This cell is a polygon formed by faces hereafter called "plaquettes". For
example, the elementary cell of the simple cubic lattice is a cube with six square plaquettes, the elementary cell of the
fcc lattice is a tetrahedron formed by four triangular plaquettes. Let $J_{i,j}$ be the interaction between two nn spins of
the plaquette. According to the definition of Toulouse \cite{Tou} the plaquette is frustrated if the parameter $P$ defined
below is negative
\begin{equation}
P=\prod_{\left<i,j\right>}\mathrm{sign}(J_{i,j}), \label{frust1}
\end{equation}
where the product is performed over all $J_{i,j}$ around the plaquette.
Two examples of frustrated plaquettes are shown in Fig. \ref{fig:IntroFP}: a triangle with three antiferromagnetic bonds
and a square with three ferromagnetic bonds and one antiferromagnetic bond.  $P$ is negative in both cases.
One sees that if one tries to put Ising spins on those plaquettes, at least one of the bonds around the
plaquette will not be satisfied. For vector spins, we show below that in the lowest energy state,
each bond is only partially satisfied.

One sees that for the triangular plaquette, the degeneracy is three, and for the square plaquette it is four, in addition
to the degeneracy associated with returning all spins.  Therefore, the degeneracy of an infinite lattice composed of such
plaquettes is infinite, in contrast to the unfrustrated case.

We emphasize that the frustration can be created with other kinds of interaction such as the Dzyaloshinski-Moriya interaction $E=-\mathbf D\cdot \left (\mathbf S_i \wedge \mathbf S_j\right)$ \cite{Dzyaloshinskii,Moriya} which favors the perpendicular spin configuration in competition with a Heisenberg exchange model which favors a collinear one.  We will return to this interaction in the section on skyrmions later in this paper.

The determination of the GS of some  frustrated spin systems as well as discussions on their properties are shown in the following.

\subsection{Non collinear spin configurations}\label{noncollinear}

Consider as examples the plaquettes shown in Fig. \ref{fig:IntroFP}.
In the case of $XY$ spins, the GS configuration is obtained
by minimizing the energy of the plaquette $E$. In the case of the triangular plaquette,
suppose that spin $\mathbf S_i$ $(i=1,2,3)$ of amplitude $S$ makes
an angle $\theta_i$ with the $\mathbf {Ox}$ axis. Writing $E$ and
minimizing it with respect to the angles $\theta_i$, one has
\begin{eqnarray}
E&=&J(\mathbf S_1\cdot \mathbf S_2+\mathbf S_2\cdot \mathbf S_3+\mathbf S_3\cdot \mathbf S_1)\nonumber \\
&=&JS^2\left[\cos (\theta_1-\theta_2)+\cos (\theta_2-\theta_3)+
\cos (\theta_3-\theta_1)\right],\nonumber \\
\frac{\partial E}{\partial \theta_1}&=&-JS^2\left[\sin
(\theta_1-\theta_2)-
\sin (\theta_3-\theta_1)\right]=0, \nonumber \\
\frac{\partial E}{\partial \theta_2}&=&-JS^2\left[\sin
(\theta_2-\theta_3)-
\sin (\theta_1-\theta_2)\right]=0, \nonumber \\
\frac{\partial E}{\partial \theta_3}&=&-JS^2\left[\sin
(\theta_3-\theta_1)- \sin (\theta_2-\theta_3)\right]=0. \nonumber
\end{eqnarray}
A solution of the last three equations is
$\theta_1-\theta_2=\theta_2 -\theta_3=\theta_3-\theta_1 =2\pi/3$.
One can also write
$$
E=J(\mathbf S_1\cdot \mathbf S_2+\mathbf S_2\cdot \mathbf
S_3+\mathbf S_3\cdot \mathbf S_1)
=-\frac{3}{2}JS^2+\frac{J}{2}(\mathbf S_1 + \mathbf S_2 + \mathbf
S_3)^2.
$$
The minimum corresponds to $\mathbf S_1 + \mathbf
S_2 + \mathbf S_3=0$ which yields the $120^\circ$
structure. This is true also for the case of
Heisenberg spin.

We can do the same calculation for the case of the frustrated
square plaquette. Suppose that the antiferromagnetic bond connects
the spins $\mathbf{S}_1$ and $\mathbf{S}_2$. We find
\begin{equation}
\theta_2-\theta_1=\theta_3 -\theta_2=\theta_4-\theta_3
=\frac{\pi}{4} \textrm { and }\theta_1-\theta_4=\frac{3\pi}{4}
\label{frust2}
\end{equation}
If the antiferromagnetic bond is equal to $-\eta J$, the solution for the angles is \cite{Berge}
\begin{equation}
\cos\theta_{32}=\cos\theta_{43}=\cos\theta_{14}\equiv \theta=\frac {1}{2}[\frac {\eta+1}{\eta}]^{1/2}\label{frust2a}
\end{equation}
and $|\theta_{21}|=3|\theta|$, where $\cos\theta_{ij}\equiv \cos\theta_{i}-\cos\theta_{j}$.
This solution exists if $| \cos\theta |\leq 1$, namely $\eta>\eta_c=1/3$. One can check that when $\eta=1$, one has $\theta
=\pi/4$, $\theta_{21}=3\pi/4$.

\begin{figure}[h!]
\centering
\includegraphics[width=2.5 in]{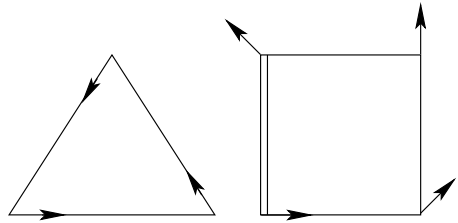}
\caption{ Non
collinear spin configuration of frustrated triangular and square
plaquettes with $XY$ spins: ferro- and antiferromagnetic
interactions
 $J$ and $-J$ are indicated by thin and double lines, respectively.}
 \label{fig:IntroNCFT}
\end{figure}

We show the GS spin configurations of the frustrated triangular and square lattices in Fig.
\ref{fig:IntroNCFT} with $XY$ spins ($N=2$).

\begin{figure}[h!]
\centering
\includegraphics[width=2.5 in]{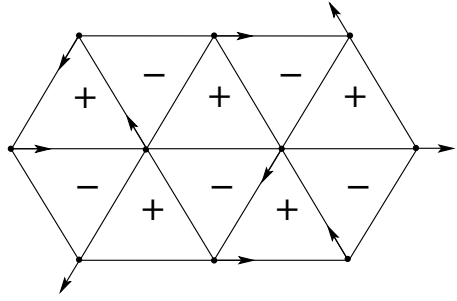}
\caption{\label{fig:IntroAFTL} Antiferromagnetic triangular
lattice with  $XY$ spins. The positive and negative chiralities
are indicated by $+$ and $-$.}
\end{figure}

At this stage, we note that the two GS found above have a two-fold degeneracy  resulting from the equivalence of clockwise or counter-clockwise turning angle (noted by $+$ and $-$ in Fig. \ref{fig:IntroAFTL}) between adjacent spins on a plaquette in Fig. \ref{fig:IntroNCFT}. Therefore the
symmetry of these plaquettes is of Ising type O(1), in addition to
the symmetry SO(2) due to the invariance by global rotation of the
spins in the  plane.

It is expected from the GS symmetry of these systems that the
transitions due to the respective breaking of O(1) and SO(2)
symmetries, if they occur at different temperatures, belong
respectively to the 2D Ising universality class and to the
Kosterlitz-Thouless universality class \cite{Zinn}. The question of whether
the two phase transitions would occur at the same temperature and
the nature of their universality remains at present an open
question.

\begin{figure}[h!]
\centering
\includegraphics[width=2 in]{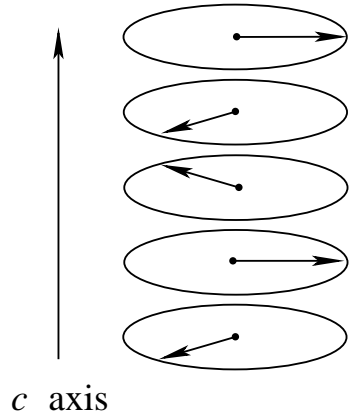}
\caption{Helical
configuration when $\varepsilon = |J_2|/J_1>\varepsilon_c= 1/4$
($J_1>0$, $J_2<0$).}
 \label{fig:IntroHC}
\end{figure}

Another example is the case of helimagnets. Consider a chain of Heisenberg spins with
ferromagnetic interaction $J_1(>0)$ between nn and
antiferromagnetic interaction
 $J_2 (<0)$ between nnn. When
$\varepsilon = |J_2|/J_1$ is larger than a critical value
$\varepsilon_c$, the spin configuration of the GS becomes non
collinear. One shows that the helical configuration displayed in
Fig. \ref{fig:IntroHC} is obtained by minimizing the interaction
energy:
\begin{eqnarray}
E&=&-J_1\sum_i\mathbf S_i\cdot\mathbf S_{i+1}
+|J_2|\sum_i\mathbf S_i\cdot\mathbf S_{i+2}\nonumber \\
&=&S^2\left[ -J_1\cos \theta+|J_2|\cos (2\theta)\right]\sum_i1\nonumber \\
\frac{\partial E}{\partial \theta}&=&S^2\left[J_1\sin \theta
-2|J_2|\sin(2\theta)\right]\sum_i1=0\nonumber \\
&=&S^2\left[J_1\sin \theta -4|J_2|\sin\theta \cos \theta
\right]\sum_i1=0,
\end{eqnarray}
where one has supposed that the angle between nn spins is
$\theta$.

The two solutions are
$$
\sin \theta=0 \longrightarrow \theta=0 \hspace{0.2cm}
\textrm{(ferromagnetic solution)}
$$
and
\begin{equation}
\cos \theta=\frac{J_1}{4|J_2|} \longrightarrow \theta= \pm \arccos
\left(\frac{J_1}{4|J_2|}\right).
\end{equation}
The last solution is possible if $-1\le \cos\theta \le 1$, i.e.
$J_1/\left(4|J_2|\right)\le 1$ or $|J_2|/J_1\ge 1/4 \equiv
\varepsilon_c$.

This is shown in Fig. \ref{fig:IntroHC}.  There are two degenerate configurations corresponding to clockwise and counter-clockwise turning angles as in the previous examples.

Let us enumerate two frequently encountered frustrated spin
systems where the nn interaction is antiferromagnetic: the fcc
lattice and the hcp lattice. These two lattices are formed by
stacking tetrahedra with four triangular faces.  The frustration
due to the lattice structure such as in these cases is called
"geometry frustration" \cite{DiepFSS}.

\section{Exactly solved frustrated models}

\begin{figure}[h!]
\centering
\includegraphics[width=2 in]{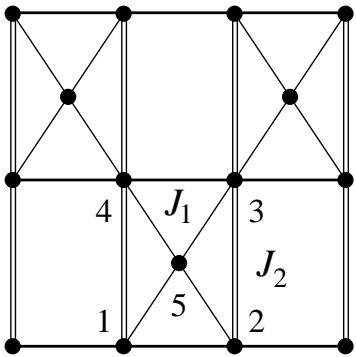}
\caption{\label{modelsK}Kagom\'{e} lattice: Interactions between
nearest neighbors and between next-nearest neighbors, $J_{1}$ (horizontal  and diagonal bonds) and
$J_{2}$ (vertical bonds), are shown by single and double bonds, respectively.
}
\end{figure}

\begin{figure}[h!]
\centering
\includegraphics[width=2 in]{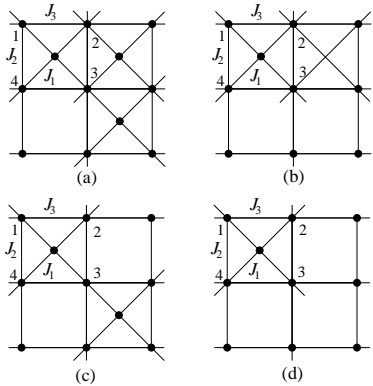}
\caption{\label{modelsDS}
Elementary cells of periodically dilute
centered square lattice: (a)  three-center case, (b)
two-adjacent-center case, (c) two-diagonal-center case, (d)
one-center case.  Interactions along diagonal, vertical and
horizontal bonds are $J_{1}$, $J_{2}$, and $J_{3}$, respectively.
}
\end{figure}

\begin{figure}[h!]
\centering
\includegraphics[width=2 in]{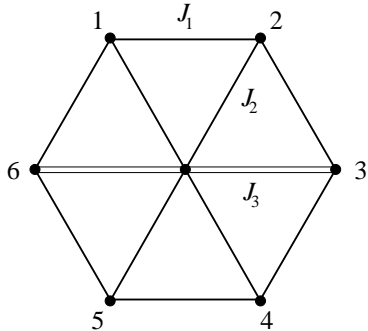}
\caption{\label{modelsHC}
Unit cell of the centered honeycomb
lattice: heavy, light, and double-light bonds denote the
interactions $J_{1}$, $J_{2}$, and $J_{3}$, respectively. The
sites on the honeycomb are numbered from 1 to 6 for decimation
demonstration (see text).
}
\end{figure}

The 2D Ising model with non-crossing interactions is exactly soluble. Instead of finding the partition function one can map the model on a 16-vertex model or a 32-vertex model. The resulting vertex model will be exactly soluble. We have applied this
method for finding the exact solution of several Ising frustrated models in 2D lattices with non-crossing interactions shown in Figs. \ref{modelsK}-\ref{modelsHC}.

 Details have been given in Ref. \cite{Giacomini}.  I outline here only a simplified formulation of a model for illustration. The aim is to discuss the results.  As we will see these models possess spectacular phenomena due to the frustration.

We take the case of the centered honeycomb lattice. The Hamiltonian of this model is as follows :

\begin{equation}
H=-J_{1}\sum_{(ij)} \sigma_{i}\sigma_{j}-J_{2}\sum_{(ij)}
\sigma_{i}\sigma_{j}-J_{3}
\sum_{(ij)}\sigma_{i}\sigma_{j}\label{eq26}
\end{equation}
where $\sigma_{i}=\pm 1$  is an Ising spin occupying the lattice
site i , and the first, second, and third sums run over the spin
pairs connected by heavy, light, and doubly light bonds,
respectively (see Fig. \ref{modelsHC}). When $J_{2}=J_{3}=0$, one
recovers the honeycomb lattice, and when $J_{1}=J_{2}=J_{3}$,
one has the triangular lattice.

Let us denote the central spin in a lattice cell, shown in Fig.
\ref{modelsHC}, by $\sigma$, and number the other spins from
$\sigma_{1}$ to $\sigma_{6}$. The Boltzmann weight associated to
the elementary cell is given by

\begin{eqnarray*}
W = \exp[K_{1}(\sigma_{1}\sigma_{2}+
\sigma_{2}\sigma_{3}+\sigma_{3}\sigma_{4}+
\sigma_{4}\sigma_{5}+\sigma_{5}\sigma_{6}+
\sigma_{6}\sigma_{1})+
\end{eqnarray*}
\begin{equation}
K_{2}\sigma(\sigma_{1}+\sigma_{2}+\sigma_{4}+\sigma_{5})+
K_{3}\sigma(\sigma_{3}+\sigma_{6})]\label{eq27}
\end{equation}

The partition function is written as

\begin{equation}
Z=\sum_{\sigma}\prod_{c} W\label{eq28}
\end{equation}
where the sum is performed over all spin configurations and the
product is taken over all elementary cells of the lattice.
Periodic boundary conditions are imposed. Since there is no
crossing-bond interaction, the model is exactly soluble. To obtain
the exact solution, we decimate the central spin of each
elementary cell of the lattice. In doing so, we obtain a honeycomb
Ising model with multispin interactions.

After decimation of each central spin, the Boltzmann factor
associated to an elementary cell is given by

\begin{eqnarray*}
W' = 2\exp[K_{1}(\sigma_{1}\sigma_{2}+
\sigma_{2}\sigma_{3}+\sigma_{3}\sigma_{4}+
\sigma_{4}\sigma_{5}+\sigma_{5}\sigma_{6}+
\sigma_{6}\sigma_{1})]\times
\end{eqnarray*}
\begin{equation}
\cosh [K_{2}(\sigma_{1}+\sigma_{2}+\sigma_{4}+\sigma_{5})+
K_{3}(\sigma_{3}+\sigma_{6})]\label{eq29}
\end{equation}

This model is equivalent to a
special case of the 32-vertex model on the triangular lattice that
satisfies the free-fermion condition as seen in the following.

\begin{figure}[h!]
\centering
\includegraphics[width=2.2 in]{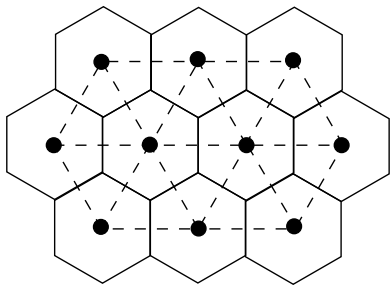}
\caption{ The honeycomb lattice and the dual
triangular lattice, with their bonds indicated by dashed lines.
\label{re-fig12}}
\end{figure}

Let us consider the dual lattice of the honeycomb lattice, i.e.
the triangular lattice \cite{Bax}. The sites of the dual lattice
are placed at the center of each elementary cell and their bonds
are perpendicular to bonds of the honeycomb lattice, as it is
shown in Fig. \ref{re-fig12}.

\begin{figure}[h!]
\centering
\includegraphics[width=2 in]{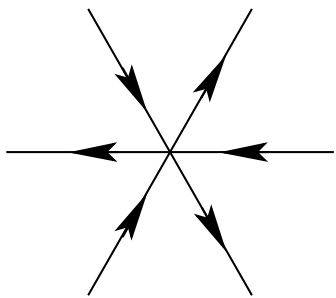}
\caption{ The standard arrow configuration for the triangular
lattice. \label{re-fig13}}
\end{figure}

Each site of the triangular lattice is surrounded by 6 sites of
the honeycomb lattice. At each bond of the triangular lattice we
associate an arrow. We take the arrow configuration shown in Fig.
\ref{re-fig13} as the standard one. We can establish a two-to-one
correspondence between spin configurations of the honeycomb
lattice and arrow configurations in the triangular lattice. This
can be done in the following way : if the spins on either side of
a bond of the triangular lattice are equal ( different ), place an
arrow on the bond pointing in the same ( opposite ) way as the
standard. If we do this for all bonds, then at each site of the
triangular lattice there must be an even number of non-standard
arrows on the six incident bonds, and hence an odd number of
incoming ( and outgoing ) arrows. This is the property that
characterizes the 32 vertex model on the triangular lattice.

\begin{figure}[h!]
\centering
\includegraphics[width=3 in]{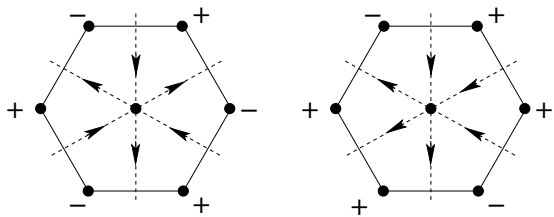}
\caption{ Two cases of the correspondence between
arrow configurations and spin configurations. \label{re-fig14}}
\end{figure}

In Fig. \ref{re-fig14} we show two cases of the relation between
arrow configurations on the triangular lattice and spin
configurations on the honeycomb lattice.

In consequence, the Boltzmann weights of the 32-vertex model will
be a function of the Boltzmann weights $W'(\sigma_{1},\sigma_{2},
\sigma_{3},\sigma_{4},\sigma_{5},\sigma_{6})$ , associated to a
face of the honeycomb lattice. By using the relation between
vertex and spin configurations described above and expression Eq.
(\ref{eq29}), we find

\begin{eqnarray}
\omega&= &W'(+,-,-,-,+,+) = 2e^{2K_{1}} \nonumber\\
\overline {\omega}&= & W'(+,+,-,+,+,-) = 2e^{-2K_{1}}\cosh (4K_{2}
-2K_{3})\nonumber\\
\omega_{56}&= & W'(+,-,+,-,+,+) = 2e^{-2K_{1}}\cosh (2K_{3})\nonumber\\
\overline {\omega}_{56}&= & W'(+,+,+,+,+,-) =
2e^{2K_{1}}\cosh (4K_{2})\nonumber\\
\omega_{15}&= & W'(+,+,+,-,+,+) = 2e^{2K_{1}}\cosh (2K_{2}+
2K_{3})\nonumber\\
\overline {\omega}_{15}&= & W'(+,-,+,+,+,-) =
2e^{-2K_{1}}\cosh (2K_{2})\nonumber\\
\omega_{46}&= & W'(+,-,+,+,+,+) = 2e^{2K_{1}}\cosh (2K_{2}+
2K_{3})\nonumber\\
\overline {\omega}_{46}&= & W'(+,+,+,-,+,-) =
2e^{-2K_{1}}\cosh (2K_{2})\nonumber\\
\omega_{13}&= & W'(+,+,+,+,-,+) = 2e^{2K_{1}}\cosh (2K_{2}+
2K_{3})\nonumber\\
\overline {\omega}_{13}&= & W'(+,-,+,-,-,-) =
2e^{-2K_{1}}\cosh (2K_{2})\nonumber\\
\omega_{24}&= & W'(+,-,-,-,-,-) = 2e^{2K_{1}}\cosh (2K_{2}+
2K_{3})\nonumber\\
\overline {\omega}_{24}&= & W'(+,+,-,+,-,+) =
2e^{-2K_{1}}\cosh (2K_{2})\nonumber\\
\omega_{14}&= & W'(+,+,+,+,+,+) = 2e^{6K_{1}}\cosh (4K_{2}+
2K_{3})\nonumber\\
\overline {\omega}_{14}&= & W'(+,-,+,-,+,-) = 2e^{-6K_{1}}\nonumber\\
\omega_{23}&= & W'(+,-,-,-,+,-) = 2e^{-2K_{1}}\cosh (2K_{3})\nonumber\\
\overline {\omega}_{23}&= & W'(+,+,-,+,+,+) = 2e^{2K_{1}}\cosh (4K_{2})\nonumber\\
\omega_{25}&= & W'(+,-,-,+,-,-) = 2e^{-2K_{1}}\cosh (2K_{3})\nonumber\\
\overline {\omega}_{25}&= & W'(+,+,-,-,-,+) = 2e^{2K_{1}}\nonumber\\
\omega_{36}&= & W'(+,-,+,+,-,+) = 2e^{-2K_{1}}\cosh (2K_{3})\nonumber\\
\overline {\omega}_{36}&= & W'(+,+,+,-,-,-) = 2e^{2K_{1}}\nonumber\\
\overline {\omega}_{34}&= & W'(+,+,-,+,-,-) = 2e^{-2K_{1}}
\cosh (2K_{2}-2K_{3})\nonumber\\
\omega_{34}&= & W'(+,-,-,-,-,+) = 2e^{2K_{1}}\cosh (2K_{2})\nonumber\\
\overline {\omega}_{35}&= & W'(+,+,-,-,-,-) = 2e^{2K_{1}}
\cosh (2K_{3})\nonumber\\
\omega_{35}&= & W'(+,-,-,+,-,+) = 2e^{-2K_{1}}\nonumber\\
\overline {\omega}_{45}&= & W'(+,+,-,-,+,-) = 2e^{-2K_{1}}
\cosh (2K_{2}-2K_{3})\nonumber\\
\omega_{45}&= & W'(+,-,-,+,+,+) = 2e^{2K_{1}}\cosh (2K_{2})\nonumber\\
\overline {\omega}_{12}&= & W'(+,-,+,-,-,+) = 2e^{-2K_{1}}
\cosh (-2K_{2}+2K_{3})\nonumber\\
\omega_{12}&= & W'(+,+,+,+,-,-) = 2e^{2K_{1}}\cosh (2K_{2})\nonumber\\
\overline {\omega}_{26}&= & W'(+,+,+,-,-,+) = 2e^{2K_{1}}
\cosh (2K_{3})\nonumber\\
\omega_{26}&= & W'(+,-,+,+,-,-) = 2e^{-2K_{1}}\nonumber\\
\omega_{16}&= & W'(+,+,-,-,+,+) = 2e^{2K_{1}}\cosh (2K_{2})\nonumber\\
\overline {\omega}_{16}&= & W'(+,-,-,+,+,-) = 2e^{-2K_{1}} \cosh
(2K_{2}-2K_{3})\nonumber\\
&&\label{eq30}
\end{eqnarray}

Using the above expressions, the critical temperature of the model is determined from the equation (see details in Ref. \cite{Diep91a}):

\begin{eqnarray}
&&e^{2K_{1}}+e^{-2K_{1}}\cosh (4K_{2}-2K_{3})\nonumber \\
&&+ 2e^{-2K_{1}}\cosh (2K_{3})+
2e^{2K_{1}}+\nonumber \\
&&e^{6K_{1}}\cosh (4K_{2}+2K_{3})+e^{-6K_{1}}=
2\mbox{max}\{e^{2K_{1}}+\nonumber \\
&&e^{-2K_{1}}\cosh (4K_{2}-2K_{3})\:\:;\:\: \nonumber\\
&&e^{2K_{1}}+ e^{-2K_{1}}\cosh (2K_{3}); e^{6K_{1}}\cosh
(4K_{1}+2K_{3})+e^{-6K_{1}}\}\nonumber \\
&&\label{eq31}
\end{eqnarray}
The solutions of this equation are given below for some special cases.

Following the case studied above, we can study other
2D Ising models without crossing bonds shown in Figs. \ref{modelsK}-\ref{modelsDS}:
after decimation of the central spin in each square, these
models can be mapped into a special case of the 16-vertex
model which yields the exact solution for the critical surface (see details in Ref. \cite{Giacomini}).

Before showing some results in the  space of interaction parameters, let us introduce the definitions of disorder line and reentrant phase.

\subsection{Disorder line, reentrance}
 It is not the purpose of this review to enter technical details. I would rather like to describe the physical meaning of the disorder line and the reentrance. A full technical review has been given in Ref. \cite{Giacomini}.

 Disorder solutions are very useful for
clarifying the phase diagrams of anisotropic models and also imply
constraints on the analytical behavior of the partition function
of these models.

A great variety of anisotropic models (with different
coupling constants in the different directions of the
lattice) are known to posses remarkable sub-manifolds in the
space of parameters where the partition function is
computable and takes a very simple form. These are the
disorder solutions.

All the methods applied for obtaining these solutions rely
on the same mechanism : a certain local decoupling of the
degrees of freedom of the model results in an
effective reduction of dimensionality for the lattice
system. Such a property is provided by a simple local
condition imposed on the Boltzmann weights of the elementary
cell generating the lattice \cite{Ste}.  On a disorder line, a 2D system can behave as a 1D one: the dimension reduction is due to the decoupling of a degree of freedom in one direction, for instance.

This is very important while interpreting the system behavior: on one side of
the disorder line, pre-ordering fluctuations have correlation different from those of the other side.  Crossing the line, the system pre-ordering correlation changes.  The dimension reduction is often necessary to realize this.

Disorder solutions have recently found interesting applications,
for example in the problem of cellular automata (for a review
see Rujan \cite{Rujan}). Moreover, they also serve to built a new
kind  of series expansion for lattice spin systems \cite{Mail}.

Let us give now a definition for the reentrance. A reentrant phase lies between two ordered phases. For example, at low temperature ($T$) the system is in an ordered phase I. Increasing $T$, it undergoes a transition to a paramagnetic phase $R$, but if one increases further $T$, the system enters another ordered phase II before becoming disordered at higher $T$.  Phase $R$ is thus between two ordered phases I and II. It is called "reentrant paramagnetic phase" or "reentrant phase".

How physically is it possible? At a first sight, it cannot be possible because the entropy of an ordered phase is smaller than that of an disordered phase so that the disordered phase $R$ cannot exist at lower $T$ than the ordered phase II.  In reality, as we will see below, phase II has always a partial disorder which compensates the loss of entropy while going from $R$ to II. The principle that entropy increases with $T$ is thus not violated.

\subsection{Phase diagram}
\subsubsection{Kagom\'{e} lattice}

A model of great interest is the Kagom\'{e}
lattice shown in Fig. \ref{modelsK}. The
Kagom\'{e} Ising lattice with nn interaction $J_{\rm 1}$ has been
solved a long time ago\cite{Ka/Na} showing no phase transition at
finite $T$ when $J_{1}$ is antiferromagnetic. Taking into account
the nnn interaction $J_{2}$, we have solved \cite{Aza87} this model by
transforming it into a 16-vertex model which satisfies the
free-fermion condition.  The critical surface is given by
\begin{eqnarray}
\frac {1}{2}\ [\exp(2K_{1}+2K_{2})\cosh (4K_{1})+
\exp(-2K_{1}-2K_{2})]&+& \nonumber\\
\cosh (2K_{1}-2K_{2})+2\cosh (2K_{1})=
2\mbox{max}\{\frac {1}{2}\ [\exp(2K_{1}&+&2K_{2})\cosh (4K_{1})+ \nonumber\\
\exp(-2K_{1}-2K_{2})]\:;\:\cosh (2K_{2}-2K_{1})&;&\cosh
(2K_{1})\}\label{eq25}
\end{eqnarray}

\begin{figure}[h!]
\centering
\includegraphics[width=3.2 in]{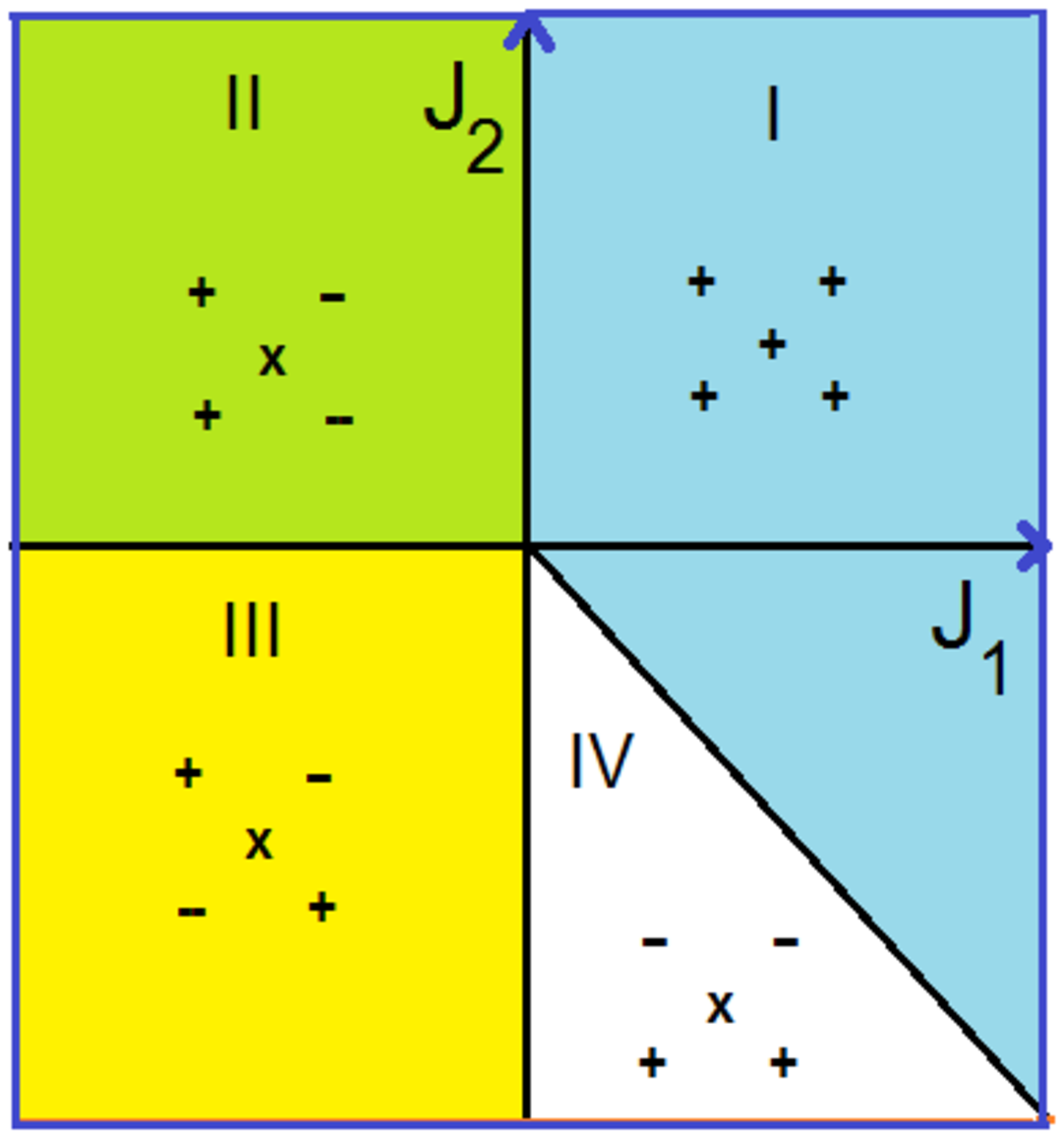}
\includegraphics[width=3.2 in]{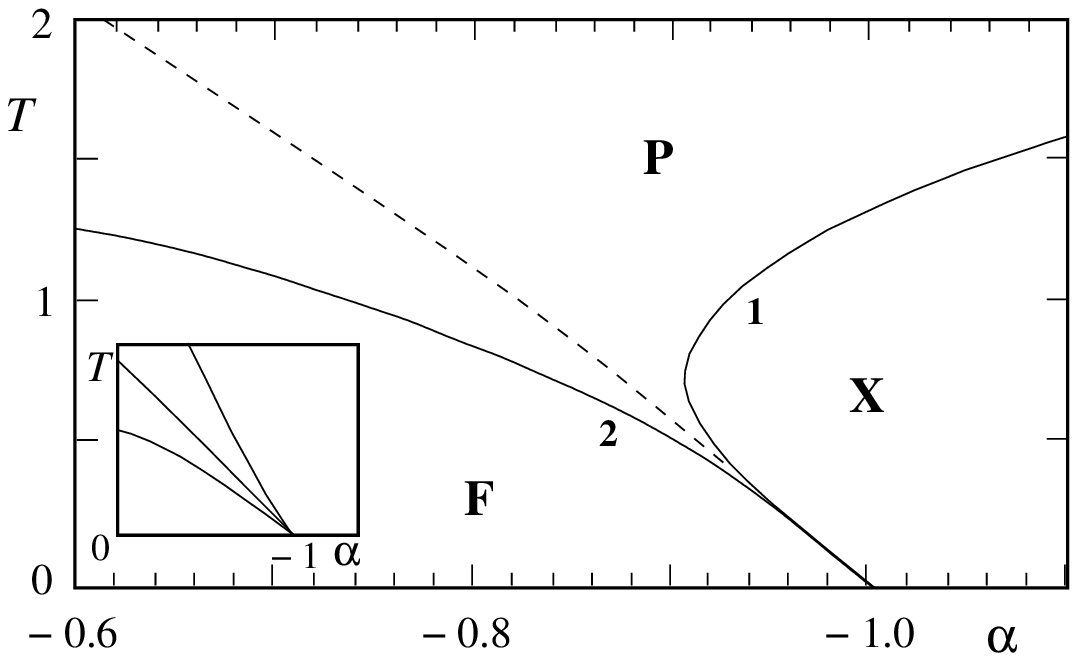}
\vspace*{8pt} \caption{Left: Ground-state phase diagram in the space ($J_1,J_2)$ where $+$, $-$ and $x$ denote up, down and free (undetermined) spins, respectively. Right: Phase diagram of the Kagom\'{e} lattice
with nnn interaction in the region $J_{1} > 0$ of the space
($\alpha=J_{2}/J_{1}, T$). $T$ is measured in the unit of
$J_{1}/k_{B}$.  Solid lines are critical lines, dashed line is the
disorder line. P, F and X stand for paramagnetic, ferromagnetic
and partially disordered phases, respectively.  The inset shows
schematically enlarged region of the endpoint. \label{re-fig20}}
\end{figure}

For the
whole phase diagram, the reader is referred to Ref. \cite{Aza87}.
We show in Fig. \ref{re-fig20} (bottom) only the small
region of $J_{2}/J_{1}$ in the phase diagram which has the
reentrant paramagnetic phase and a disorder line. This region lies around the phase boundary between two phases IV (partially disordered) and I (ferromagnetic) in Fig. \ref{re-fig20} (top).

The phase X indicates a partially ordered phase where the central spins are free. The nature of
ordering was determined by Monte Carlo (MC) simulations \cite{Aza87}.

Here
again, the reentrant phase takes place between a low-$T$ ordered
phase and a partially disordered phase. This suggests that a
partial disorder in the high-$T$ phase is necessary to ensure that
the entropy is larger than that of the reentrant phase.


When all the interactions are different in the model shown in Fig.
\ref{modelsK}, i.e. the horizontal bonds $J_{3}$, the vertical
bonds $J_{2}$ and the diagonal ones are not equal, the phase diagram becomes complicated with new
features \cite{Diep91b}:  in particular, we show that the
reentrance can occur in an {\it infinite region} of phase space.
In addition, there may be {\it several reentrant phases} occurring
for a given set of interactions when $T$ varies.

The Hamiltonian is written as
\begin{equation}
H=-J_{1}\sum_{(ij)} \sigma_{i}\sigma_{j}-J_{2}\sum_{(ij)}
\sigma_{i}\sigma_{j}-J_{3}
\sum_{(ij)}\sigma_{i}\sigma_{j}\label{eq46}
\end{equation}
where $\sigma_{i}=\pm 1$  is an Ising spin occupying the lattice site i ,
and the
first, second, and third sums run over the spin pairs
connected by
diagonal, vertical and horizontal bonds, respectively.
When
$J_{2}=0$ and $J_{1}=J_{3}$, one recovers the original nn Kagom\'{e}
lattice \cite{Ka/Na}.  The effect of $J_{2}$ in the case $J_{1}=J_{3}$
has been shown above.
\begin{figure}[h!]
\centering
\includegraphics[width=4 in]{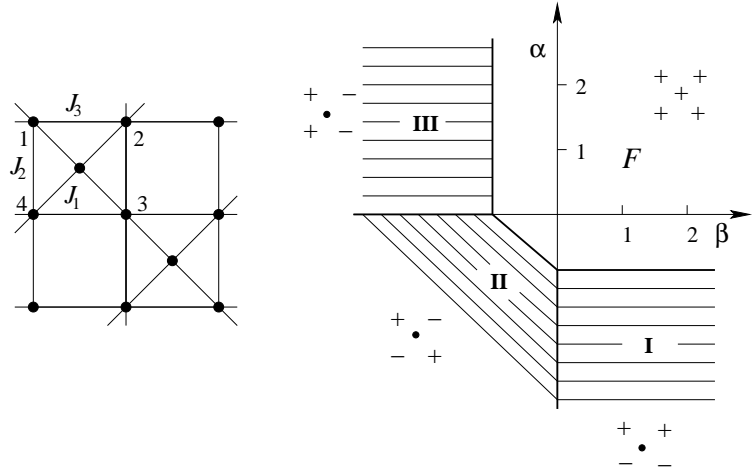}
\vspace*{8pt} \caption{ Left: Generalized Kagom\'{e} lattice:
diagonal, vertical and horizontal  bonds denote the interactions
$J_{1}$, $J_{2}$ and $J_{3}$, respectively. Right: Phase diagram
of the ground state shown in the plane ($\alpha =J_{2}/J_{1},
\beta = J_{3}/J_{1}$). Heavy lines separate different phases and
spin configuration of each phase is indicated (up, down and free
spins are denoted by +, - and o, respectively).  The three kinds
of partially disordered phases and the ferromagnetic phase are
denoted by I, II , III and F, respectively. \label{re-fig21}}
\end{figure}

The phase diagram at temperature $T=0$ is shown in Fig.
\ref{re-fig21} in the space ($\alpha = J_{2}/J_{1}$, $\beta =
J_{3}/J_{1}$) for positive $J_{1}$. The ground- state  spin
configurations are also displayed. The hatched regions indicate
the three partially disordered phases (I, II, and III) where the
central spins are free.  Note that the phase diagram is
mirror-symmetric with respect to the change of the sign of
$J_{1}$. With negative $J_{1}$ , it suffices to reverse the
central spin in the spin configuration shown in Fig.
\ref{re-fig21}. Furthermore, the interchange of $J_{2}$ and
$J_{3}$ leaves the system invariant, since it is equivalent to a
$\pi /2$ rotation of the lattice. Let us consider the effect of
the temperature on the phase diagram shown in Fig. \ref{re-fig21}.
Partial disorder in the ground state often gives rise to the
reentrance phenomenon as in systems shown above.  Therefore,
similar effects are to be expected in the present system.  As it
will be shown below, we find a new and richer behavior of the
phase diagram:  in particular,  the reentrance region is found to
be extended to infinity, unlike systems previously studied, and
for some given set of interactions, there exist {\it two disorder
lines} \index{disorder lines} which divide the paramagnetic phase
into regions of different kinds of fluctuations with a reentrant
behavior.

Following the decimation method \cite{Giacomini}, one obtains a
checkerboard Ising model with multispin interactions. This
resulting model is equivalent to a symmetric 16-vertex model which
satisfies the free-fermion condition \cite{Gaff,Suzu,Wu72}. The
critical temperature of the model is given by
\begin{eqnarray}
&&\cosh (4K_{1}) \exp (2K_{2}+2K_{3}) +\exp (-2K_{2}-2K_{3}) \nonumber\\
&&=
2\cosh (2K_{3} - 2K_{2}) \pm 4\cosh (2K_{1})\label{eq47}
\end{eqnarray}
Note that Eq. (\ref{eq47}) is invariant when changing
$K_{1}\rightarrow -K_{1}$  and interchanging $K_{2}$ and $K_{3}$
as stated earlier. The phase diagram in the three-dimensional
space ($K_{1},K_{2},K_{3}$) is rather complicated to show.
Instead, we  show in the following the phase diagram in the plane
($\beta =J_{3}/J_{1},T$) for typical values of $\alpha
=J_{2}/J_{1}$.
We just show now some interesting results in the interval $0 > \alpha > -1$. In this range of $\alpha$, there are three critical lines. The
critical line separating the F and P phases and the one separating
the PD phase I from the P phase have a common horizontal asymptote
as $\beta$ tends to infinity .  They form a reentrant paramagnetic
phase between the F phase and the PD phase I for positive b
between a value $\beta_{2}$ and infinite $\beta$ (Fig. 23).
Infinite region of reentrance like this has never been found
before this model.  As $\alpha$ decreases, $\beta_{2}$ tends to
zero and the F phase is contracted.  For $\alpha<-1$, the F phase
disappears together with the reentrance.

\begin{figure}[h!]
\centering
\includegraphics[width=4 in]{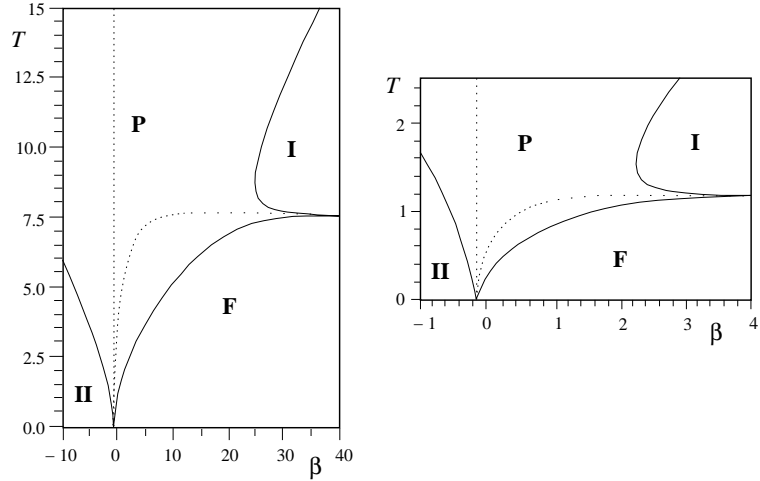}
\vspace*{8pt} \caption{ Phase diagram in the plane ($\beta
=J_{3}/J_{1},T$) for negative values of $\alpha =J_{2}/J_{1}$.
Left: $\alpha =-0.25$, Right: $\alpha=-0.8$. Solid lines are
critical lines which separate different phases: paramagnetic (P),
ferromagnetic (F), partially disordered phases of type I and II.
Dotted lines show the disorder lines. \label{re-fig23}}
\end{figure}

In the interval $0 > \alpha > -1$, the phase diagram possesses two
disorder lines (see equations in Ref. \cite{Diep91b}).
These two disorder lines are issued from a point near $\beta =-1$
for small negative $\alpha$;  this point tends to zero as $\alpha$
tends to -1 (see Fig. \ref{re-fig23}).

\subsubsection{Centered honeycomb lattice}

To obtain the exact solution of our model, we decimate the central
spin of each elementary cell of the lattice as outlined above. The resulting model is equivalent to a special case of
the 32-vertex model \cite{Sacco} on a triangular lattice that
satisfies the free-fermion condition.  The general treatment has been given in Ref. \cite{Diep91a}. Here we take a particular case when $K_{2}=K_{3}$. The critical line obtained from
Eq.(\ref{eq30}) is
\begin{eqnarray}
&&\exp(3K_{1}) \cosh (6K_{2}) + \exp(-3K_{1})\nonumber\\
&& = 3[\exp(K_{1}) +
\exp(-K_{1}) \cosh (2K_{2})]\label{eq50}
\end{eqnarray}

In the case $K_{2}=0$, the critical line is given by
\begin{eqnarray}
&&\exp(3K_{1})\cosh (2K_{3}) + \exp(-3K_{1}) \nonumber\\
&& = 3[\exp(K_{1}) +
\exp(-K_{1})\cosh (2K_{3})]\label{eq51}
\end{eqnarray}

The case $K_{3}=0$ shows on the other hand a reentrant phase.
The critical lines are
determined from the equations
\begin{equation}
\cosh (4 K_{2})=\frac {\exp(4 K_{1})+2\exp(2 K_{1})+1}{[1- \exp(4
K_{1})]\exp(2 K_{1})}\label{eq52}
\end{equation}
\begin{equation}
\cosh (4 K_{2})=\frac {3\exp(4 K_{1})+2\exp(2 K_{1})-1}{[ \exp(4
K_{1})-1]\exp(2 K_{1})}\label{eq53}
\end{equation}

\begin{figure}[h!]
\centering
\includegraphics[width=4.5 in]{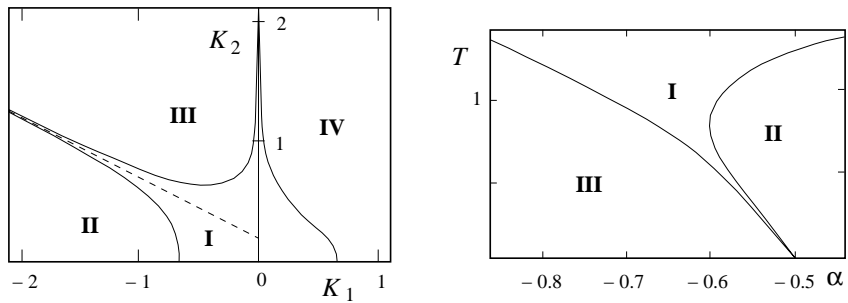}
\vspace*{8pt} \caption{ Phase diagram of the centered honeycomb
lattice with reentrance in the space ($K_{1},K_{2}$) (left) and in
the space ($T,\alpha=K_{2}/ K_{1}$) (right). I, II, III phases are
paramagnetic, partially disordered and ordered phases,
respectively.  Discontinued line is the asymptote.
\label{re-fig28}}
\end{figure}

Fig. \ref{re-fig28} shows the phase diagram obtained from Eqs.
(\ref{eq52}) and (\ref{eq53}) around the phase boundary $\alpha=-0.5$.  The reentrant paramagnetic phase
goes down to zero temperature at the boundary $\alpha=-0.5$ separating GS phases II and III
(see Fig. \ref{re-fig28} right).

Note that the honeycomb model that we have studied here does not
present a disorder solution with a
dimensional reduction.

\subsubsection{Periodically dilute centered square lattices}

In this paragraph, we show the exact results on
several periodically dilute centered square
Ising lattices by transforming them into 8-vertex models
of {\it different vertex statistical weights} that satisfy the
free-fermion condition. The dilution is introduced by taking away a
number of centered spins in a periodic manner. For a
given set of interactions, there may be five transitions with decreasing temperature with
two reentrant paramagnetic phases.   These two phases extend to
infinity in the space of interaction parameters. Moreover,  two
additional reentrant phases are found, each in a limited region
of phase space \cite{Diep92}.

Let us consider several {\bf periodically  dilute centered square
lattices}  defined from the centered square lattice shown in Fig.
\ref{modelsDS}.

The Hamiltonian of these models  is given by
\begin{equation}
H=-J_{1}\sum_{(ij)} \sigma_{i}\sigma_{j}-J_{2}\sum_{(ij)}
\sigma_{i}\sigma_{j}-J_{3}
\sum_{(ij)}\sigma_{i}\sigma_{j}\label{eq54}
\end{equation}
where $\sigma_{i}=\pm 1$  is an Ising spin occupying the lattice
site i , and the first, second and third sums run over the spin
pairs connected by diagonal, vertical and horizontal bonds,
respectively.  All these models have at least one partially
disordered phase in the ground state, caused by the competing
interactions.

In each dilute square model shown in Fig.
\ref{modelsDS},  the reentrance occurs
along most of the critical lines when the temperature is switched
on. This is a very special feature of the models  which has not been found in other models.

\begin{figure}[h!]
\centering
\includegraphics[width=3.5 in]{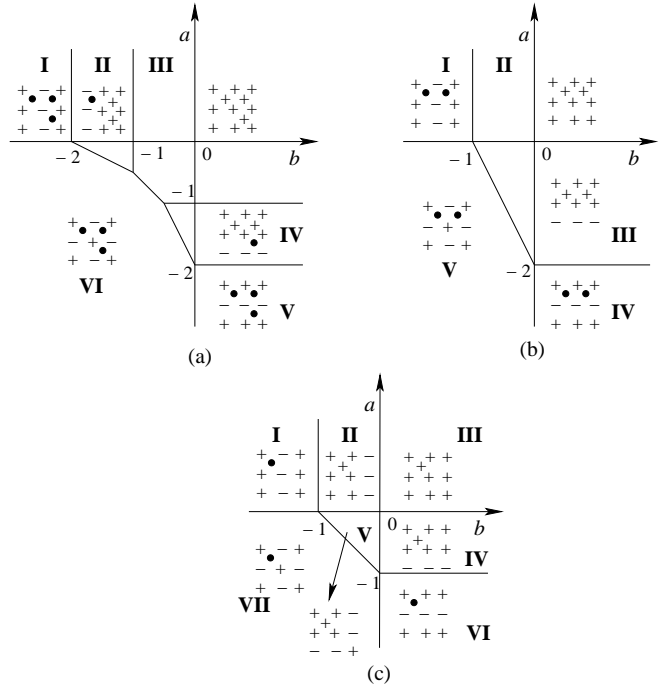}
\vspace*{8pt} \caption{ Phase diagrams in the plane ($a =J_{2}
/J_{1}$, $b =J_{3}/J_{1}$) at $T=0$ are shown for the three-center
case (a), two-adjacent center case (b), and one-center case (c).
Critical lines are drawn by heavy lines. Each phase is numbered
and the spin configuration is  indicated (+, -, and o are up,
down, and free spins, respectively). Degenerate configurations are
obtained by reversing all spins. \label{re-fig30}}
\end{figure}

Let us show in Fig. \ref{re-fig30} the phase diagrams, at $T=0$,
of the models shown in Figs. \ref{modelsDS}a, \ref{modelsDS}b and \ref{modelsDS}d, in the space ( $a,
b$ ) where $a =J_{2} /J_{1}$ and $b =J_{3}/J_{1}$. The spin
configurations in different phases are also displayed.  The
three-center case (Fig. \ref{re-fig30}a), has six phases (numbered
from I to VI), five of which (I, II, IV, V and VI) are partially
disordered (with, at least, one centered spin being free), while
the two-center case (Fig. \ref{re-fig30}b) has five phases, three
of which (I, IV, and V) are partially disordered.    Finally, the
one-center case has seven phases with three partially disordered
ones (I, VI and VII).

\begin{figure}[h!]
\centering
\includegraphics[width=4 in]{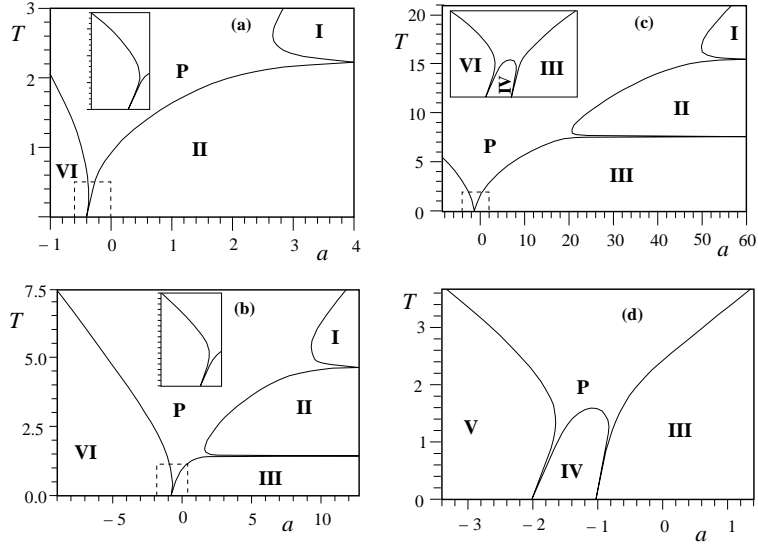}
\vspace*{8pt} \caption{ Phase diagrams in the plane ($T, a =J_{2}
/J_{1}$) for several values of $b =J_{3}/J_{1}$:(a) $b=-1.25$, (b)
$b=-0.75$, (c) $b=-0.25$, (d) $b=0.75$.  Reentrant regions on
negative sides of $a$ (limited by discontinued lines) are
schematically enlarged in the insets.  The nature of ordering in
each phase is indicated by a number which is referred to the
corresponding spin configuration in Fig. \ref{re-fig30}a. P is
paramagnetic phase. \label{re-fig31}}
\end{figure}

Without showing the detailed calculation, let us describe in Fig. \ref{re-fig31} the phase diagram of the three-center model (see Fig. \ref{modelsDS}a)
in
the space ( $a =J_{2} /J_{1}$, $T$) for typical values of $b
=J_{3}/J_{1}$.

For $b < -1$, there are two reentrances.  Fig. \ref{re-fig31}a
shows the case of $b = -1.25$ where the nature of the ordering in
each phase is indicated using the same numbers of corresponding
ground state configurations (see Fig. \ref{re-fig30}). Note that
all phases (I, II and VI) are partially disordered:  the centered
spins which are disordered at $T=0$ (Fig. \ref{re-fig30}a) remain
so at all $T$. As seen, one paramagnetic reentrance is found in a
small region of negative $a$ (schematically enlarged in the inset
of Fig. \ref{re-fig31}a), and the other on the positive $a$
extending to infinity.  The two critical lines in this region have
a common horizontal asymptote.

For $-1 < b < - 0.5$, there are three reentrant paramagnetic
regions as shown in Fig. \ref{re-fig31}b:  the reentrant region on
the negative $a$ is very narrow (inset), and the two on the
positive $a$ become so narrower while $a$ goes to infinity that
they cannot be seen on the scale of Fig. \ref{re-fig31}. Note that
the critical lines in these regions have horizontal asymptotes.
For a large value of $a$, one has five transitions with decreasing
$T$: paramagnetic state - partially disordered phase I - reentrant
paramagnetic phase - partially disordered phase II - reentrant
paramagnetic phase- ferromagnetic phase (see Fig. \ref{re-fig31}b
). So far, this is the first model that exhibits such successive
phase transitions with two reentrances.

For $- 0.5 < b < 0$, there is an additional reentrance for $a <
-1$:  this is shown in the inset of Fig. \ref{re-fig31}c. As $b$
increases from negative values, the ferromagnetic region (III) in
the phase diagram "pushes" the two partially disordered phases (I
and II) toward higher $T$.

At $b = 0$, these two  phases
disappear at infinite $T$, leaving only the ferromagnetic phase.

For positive b, there are thus only two reentrances remaining on a
negative region of $a$, with endpoints at $a = -2$ and $a = -1$,
at $T = 0$ (see Fig. \ref{re-fig31}d).

In conclusion, we summarize that in  dilute square lattice models shown in Fig. \ref{modelsDS}a, we have found
two reentrant phases occurring on the temperature scale at a given
set of interaction parameters. A striking feature is the existence
of a reentrant phase between {\it two partially disordered phases}
which has not been found so far in any other model (we recall that
in other models, a reentrant phase is found between an ordered
phase and a partially disordered phase).

\subsection{Summary and discussion}

The present section shows spectacular phenomena due to the frustration. What to be retained is those phenomena occur around the boundary of two phases of different ground states, namely different symmetries. These phenomena include

1) the partial disorder at equilibrium: disorder is not equally shared on all particles as usually the case in unfrustrated systems,

2) the reentrance: this occurs around the phase boundary when $T$ increases $\rightarrow$ the phase with larger entropy will win at finite $T$. In other words, this is a kind of selection by entropy.

3) the disorder line: this line occurs in the paramagnetic phase. It separates the pre-ordering zones between two nearby ordered phases.

The partial disorder and the reentrance  which occur in exactly solved Ising
systems shown above are expected to occur also in models other than the Ising one as well as in some three-dimensional
systems. Unfortunately, these systems cannot be exactly solved. One has to use approximations or numerical simulations to
study them. This renders difficult the interpretation of the results. Nevertheless, in the light of what has been found in
exactly solved systems, we can introduce the necessary ingredients into the model under study if we expect the same
phenomenon to occur.

As seen above, the most important ingredient for a partial disorder and a reentrance to occur at low $T$ in the Ising model
is the existence of a number of free spins in the ground state.

In three dimensions, apart from a particular exactly solved case \cite{Horiguchi} showing a reentrance,  a few Ising systems
such as the fully frustrated simple cubic lattice \cite{Blan,Diep85b}, a stacked triangular Ising
antiferromagnet \cite{Blan85,Nagai} and a body-centered cubic (bcc) crystal \cite{Aza89b} exhibit a partially disordered
phase in the ground state. We believe that reentrance should also exist in the phase space of such systems though evidence
is found numerically only for the bcc case \cite{Aza89b}.

In two dimensions, a few non-Ising models show also evidence of a reentrance. For the $q$-state Potts model,\index{Potts
model} evidence of a reentrance is found in a study of the two-dimensional frustrated Villain lattice (the so-called
piled-up domino model) by a numerical transfer matrix calculation \cite{foster1,foster2}. It is noted
that the reentrance occurs near the fully frustrated situation, i.e. $\alpha_c=J_{AF}/J_F=-1$ (equal antiferromagnetic and
ferromagnetic bond strengths), for $q$ between $\simeq 1.0$ and $\simeq 4$.  Note that there is no reentrance in the case
$q=2$. Below (above) this $q$ value, the reentrance occurs above (below) the fully frustrated point $\alpha_c$. For $q$ larger than $\simeq 4$, the reentrance disappears \cite{foster2}.

The necessary condition for the occurrence of a partial disorder at finite $T$ is thus the
existence of several kinds of site with different energies in the ground state. This has been so far verified in a number
of systems as shown above.

\section{Physics of thin films: surface magnetism, background}\label{thinfilms}

\subsection{Surface parameters}
Surface physics has been rapidly developed in the last 30 years thanks to the progress in the fabrication and the characterization of films of very thin thickness down to a single atomic layer.  A lot of industrial applications have been made in memory storage, magnetic censors, ... using properties of thin films.

Theory and simulation have also been in parallel developed to understand these new properties and to predict further interesting effects.  In the following I introduce some useful microscopic mechanisms which help understand macroscopic effects observed in experiments.

The existence of a surface on a crystal causes a lot of modifications at the microscopic levels. First, the lack of neighbors of atoms on the surface causes modifications in their electronic structure giving rise to modifications in electron orbital and atom magnetic moment by for example the spin-orbit coupling and in interaction parameters with neighboring atoms (exchange interaction, for example). In addition, surfaces can have impurities, defects (vacancies, islands, dislocations, ...). In short, we expect that the surface parameters are not the same as the bulk ones.  As a consequence, we expect physical properties at and near a surface are different from those in the bulk.  For the fundamental theory of magnetism and its application to surface physics, the reader is referred to Ref. \cite{DiepTM}.

In the following we outline some principal microscopic mechanisms which dominate properties of thin films.

\subsection{Surface spin-waves: simple examples}

In magnetically ordered systems, spin-wave (SW) excitations dominate thermodynamic properties at low $T$.  The presence of a surface modifies the SW spectrum.  We show below that it gives rise to SW modes localized near the surface. These modes lie outside the bulk SW spectrum and modify the low-$T$ behavior of thin films.

Let us calculate these modes in some simple cases. We give below for pedagogical purpose some technical details.

We consider a thin film of $N_T$ layers stacked in the $z$ direction.
The Hamiltonian is written as

\begin{eqnarray}
{\cal{H}}&=&-2\sum_{<i,j>}J_{ij} \mathbf S_i\cdot \mathbf S_j
-2\sum_{<i,j>}D_{ij} S_i^z S_j^z \nonumber\\
&=&  -2\sum_{\langle i,j\rangle}J_{ij} \left(
S_i^zS_j^z+\frac{1}{2}(S_i^+S_j^-+S_i^-S_j^+)\right)
-2\sum_{<i,j>} D_{ij}S_i^z S_j^z \nonumber\\
&& \label{surf30}
\end{eqnarray}
where $J_{ij}$ is the exchange interaction between to nn Heisenberg quantum spins, and $D_{ij}>0$ denotes  an exchange anisotropy. $S_i^+$ and $S_j^-$ are the standard spin operators $S_j^{\pm}= S_j^{x}\pm iS_j^{y}$.

For simplicity, we suppose no defects and impurities at the surface and all interactions are identical for surface and bulk spins. The microscopic mechanism which governs thermodynamic properties of magnetic materials at low temperatures is the  spin waves.
The presence of a surface often causes spin-wave modes localized at and near the surface. These modes cause in turn a diminution of the surface magnetization and the magnetic transition temperature. There are several methods to calculate the spin-wave spectrum such as (see examples given in Ref. \cite{DiepTM}) the method of equation of motion, the Holstein-Primakoff method and  the Green's function method using a correlation function between two spin operators. Here we use for illustration the Green's function method which the author has developed for thin films (see details in Ref. \cite{Diep1979,DiepTF91}).  This method shall be generalized below for helimagnets and other systems with non-collinear spin configurations.

We define the following double-time Green's function

\begin{equation}\label{gfd}
G_{i,j}(t,t')=\langle\langle S_i^+(t);S_j^-(t')\rangle\rangle
\end{equation}
The equation of motion of  $G_{i,j}(t,t')$ is written as

\begin{equation}\label{eqmvt}
i\hbar \frac{dG_{i,j}(t,t')}{dt}=(2\pi)^{-1}\langle[S_i^+(t),S_j^-(t')]\rangle +
\langle\langle [S_i^+;{\cal H}](t);S_j^-(t')\rangle\rangle
\end{equation}
\noindent where $[...]$ is the boson commutator and
$\langle...\rangle$ the thermal average in the canonical ensemble defined as
\begin{equation}
\langle F\rangle=\mbox{Tr}\mbox{e}^{-\beta {\cal{H}}}F/\mbox{Tr}\mbox{e}^{-\beta {\cal{H}}}
\end{equation}
\noindent with $\beta=1/k_BT$.
The commutator of the right-hand side of Eq. (\ref{eqmvt})
 generates functions of higher orders.  In the first
 approximation, these functions can be reduced with the help of the Tyablikov
 decoupling \cite{Tyablikov} as follows

\begin{equation}
\langle\langle S_m^zS_i^+;S_j^-\rangle\rangle
\simeq \langle S_m^z\rangle\langle\langle S_i^+;S_j^-\rangle\rangle,
\end{equation}
We obtain then the same kind of Green's function defined in Eq. (\ref{gfd}).
As the system is translation-invariant in the $xy$ plane, we use the following Fourier transforms

\begin{equation}
G_{i,j}(t,t')=\frac{1}{\Delta} \int \int d{\vec k_{xy}} \frac{1}{2
\pi}\int^{+\infty}_{-\infty} d\omega\,\mbox{e}^{-i\omega( t-t')} \,
g_{n,n'}(\omega,\vec k_{xy})\, \mbox{e}^{i\vec k_{xy}.(\vec R_i-
\vec R_j)}
\end{equation}
where  $\omega$ is the SW (magnon) pulsation (frequency), $\vec k_{xy}$ the wave vector
parallel to the surface,  $\vec R_i$ the position
of the spin at the site $i$, $n$ and $n'$ are respectively the indices of the planes to
which $i$ and $j$ belong ($n=1$ is the index of the surface).
The integration on $\vec k_{xy}$ is performed within the first Brillouin zone in the $xy$ plane.
Let  $\Delta$ be the surface of that zone.
Equation  (\ref{eqmvt}) becomes

\begin{equation}\label{greenf}
(\hbar\omega-A_n) g_{n,n'}+B_n(1-\delta_{n,1})g_{n-1,n'}+
C_n(1-\delta_{n,N_T}) g_{n+1,n'}= 2\delta_{n,n'}<S_n^z>
\end{equation}
where the factors $(1-\delta_{n,1})$ and $(1-\delta_{n,N_T})$
are added to remove $C_n$ and $B_n$ terms for the first and the last layer.
The  coefficients $A_n$,  $B_n$ and  $C_n$ depend on the crystalline lattice of the film.
We give here some examples:

\begin{itemize}
\item Film of simple cubic lattice
\begin{eqnarray}
A_n&=& -2J_n<S_n^z>C\gamma_k+2C(J_n+D_n)<S_n^z> \nonumber\\
&&+2(J_{n,n+1}+D_{n,n+1})<S_{n+1}^z>\nonumber\\
&&+2(J_{n,n-1}+D_{n,n-1})<S_{n-1}^z> \\
B_n&=&2J_{n,n-1}<S_n^z> \\
C_n&=&2J_{n,n+1}<S_n^z>
\end{eqnarray}
where $C=4$ (in-plane coordination number) and $\gamma_k=\frac{1}{2}[\cos (k_xa)+\cos (k_ya)]$.

\item Film of body-centered cubic lattice
\begin{eqnarray}
A_n&=&8(J_{n,n+1}+D_{n,n+1})<S_{n+1}^z>\nonumber\\
&&+8(J_{n,n-1}+D_{n,n-1})<S_{n-1}^z> \\
B_n&=&8J_{n,n-1}<S_n^z>\gamma_k \\
C_n&=&8J_{n,n+1}<S_n^z>\gamma_k
\end{eqnarray}
where $\gamma_k=\cos (k_xa/2)\cos (k_ya/2)$
\end{itemize}

Writing Eq. (\ref{greenf}) for $n=1,2,...,N_T$, we obtain a system of $N_T$ equations which can be put in a matrix form

\begin{equation}\label{matrix}
{\bf M}(\omega){\bf g}={\bf u}
\end{equation}
where ${\bf u}$ is a column matrix whose n-th element is
$2\delta_{n,n'}<S_n^z>$.

For a given
$\vec k_{xy}$ the magnon dispersion relation
$\hbar\omega(\vec k_{xy})$ can be obtained by solving the secular equation $det |{\bf M}|=0$.
There are  $N_T$ eigenvalues $\hbar\omega_{i}$ ($i=1,...,N_T$)
for each $\vec k_{xy}$. It is obvious that $\omega_{i}$ depends
on all $\langle S_n^z\rangle$ contained in the coefficients $A_n$,
$B_n$ and $C_n$.

To calculate the thermal average of the  magnetization  of the layer
$n$  in the case where $S=\frac{1}{2}$, we use the following relation
(see chapter 6 of Ref.  \cite{DiepTM}):

\begin{equation}\label{lm}
\langle S_n^z\rangle=\frac{1}{2}-\langle S_n^-S_n^+\rangle
\end{equation}
where $\langle S_n^-S_n^+\rangle$ is given by the following spectral theorem

\begin{eqnarray}\label{fou}
   \langle {S^-_i} {S^+_j}\rangle &=&
   \lim_{\epsilon\to0}\frac{1}{\Delta}
   \int
   \int d{\vec k_{xy}}
   \int\limits_{-\infty}^{+\infty}\frac{i}{2\pi}
   \left[g_{n,n'}(\omega+i\epsilon)-
          g_{n,n'}(\omega-i\epsilon) \right] \nonumber\\
   &&\times\frac{d\omega}{e^{\beta\omega}-1}
   \mbox{e}^{i{\vec k_{xy}}.({\vec R_i}-{\vec R_j})}.
\end{eqnarray}
$\epsilon$ being an infinitesimal positive constant.
Equation  (\ref{lm}) becomes

\begin{equation}\label{lm1}
\langle S_n^z\rangle=\frac{1}{2}-
   \lim_{\epsilon\to0}\frac{1}{\Delta}
   \int
   \int d{\vec k_{xy}}
   \int\limits_{-\infty}^{+\infty}\frac{i}{2\pi}
   \left[ g_{n,n}(\omega+i\epsilon)-
          g_{n,n}(\omega-i\epsilon) \right]
\frac{d\omega}{\mbox{e}^{\beta\hbar \omega}-1}
\end{equation}
where the Green's function  $g_{n,n}$
is obtained by the solution of Eq. (\ref{matrix})

\begin{equation}\label{gnn}
g_{n,n}=\frac{|{\bf M}|_n}{|{\bf M}|}
\end{equation}
$|{\bf M}|_n $ is the determinant obtained by replacing
the n-th column of $|{\bf M}|$ by ${\bf u}$.

To simplify the notations we put $\hbar\omega_{i}=E_i$ and $\hbar\omega=E$
in the following.  By expressing
\begin{equation}
|{\bf M}|=\prod_i(E-E_{i})
\end{equation}
we see that $E_{i}$ ($i=1,...,N_T$) are the poles of the Green's function. We can therefore rewrite $g_{n,n}$ as

\begin{equation}\label{gnn1}
g_{n,n}=\sum_i\frac{f_n(E_i)}
{E-E_i}
\end{equation}
\noindent where $f_n(E_i)$ is given by
\begin{equation}\label{fn}
f_n(E_i)= \frac{|{\bf M}|_n(E_i)}
{\prod_{j\neq i}(E_i-E_{j})}
\end{equation}
Replacing Eq. (\ref{gnn1}) in Eq. (\ref{lm1}) and making use of the following identity

\begin{equation}\label{id}
\frac {1}{x-i\eta} - \frac {1}{x+i\eta}=2\pi i\delta (x)
\end{equation}
we obtain
\begin{equation}\label{lm2}
\langle S_n^z\rangle=\frac{1}{2}-
   \frac{1}{\Delta}
   \int
   \int dk_xdk_y
   \sum_{i=1}^{N_T}\frac{f_n(E_i)}
   {\mbox{e}^{\beta E_i}-1}
\end{equation}
where $n=1,...,N_T$.

As $<S_n^z>$ depends on the magnetizations of the neighboring layers via $E_i (i=1,...,N_T)$,
we should solve by iteration the equations
(\ref{lm2}) written for all layers, namely for  $n=1,...,N_T$, to obtain the layer magnetizations at a given temperature $T$.

The critical temperature $T_c$ can be calculated in a self-consistent manner by iteration, letting all  $<S_n^z>$  tend to zero.

\begin{figure}[h!]
\centering
\includegraphics[width=5 cm]{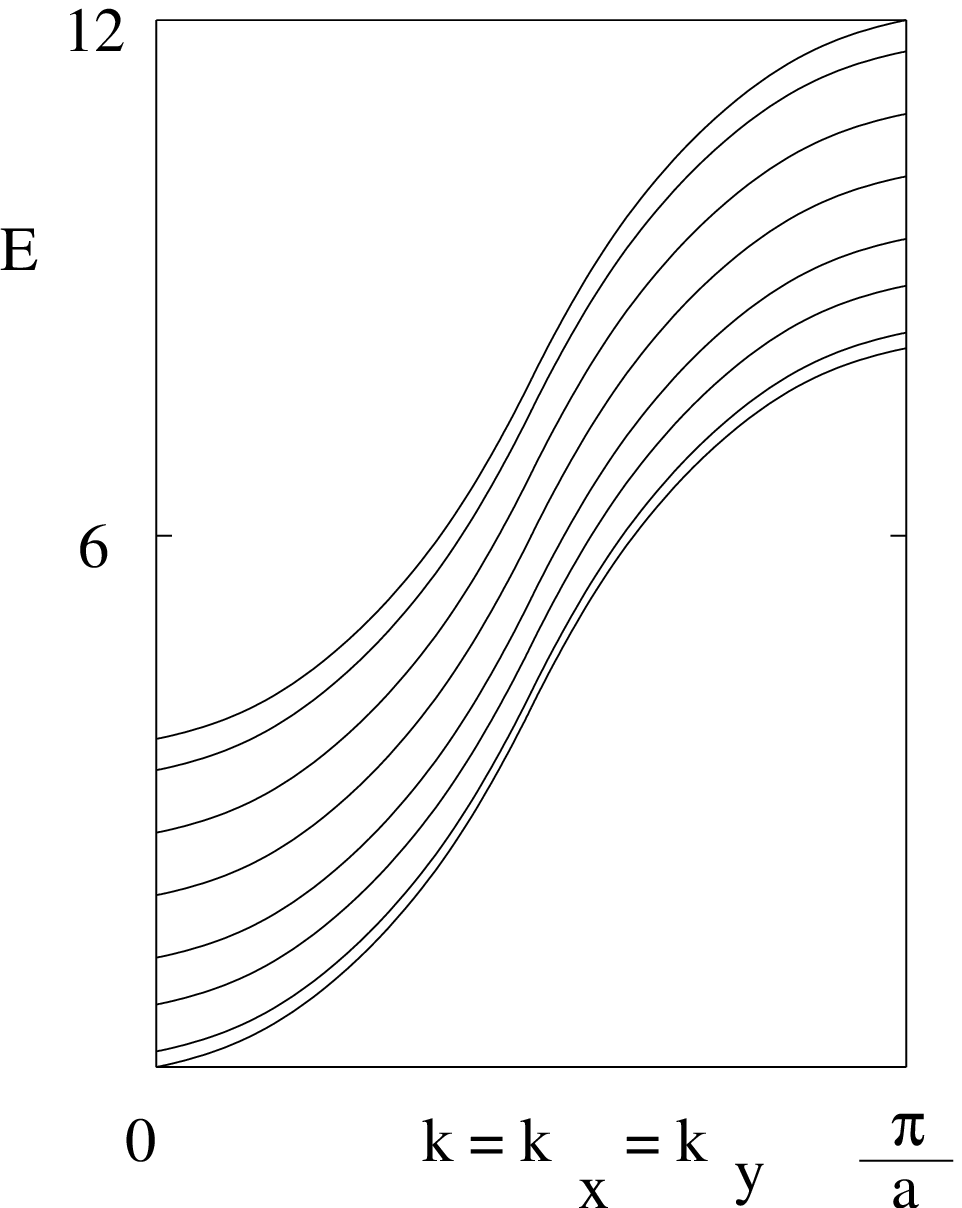}
\includegraphics[width=5 cm]{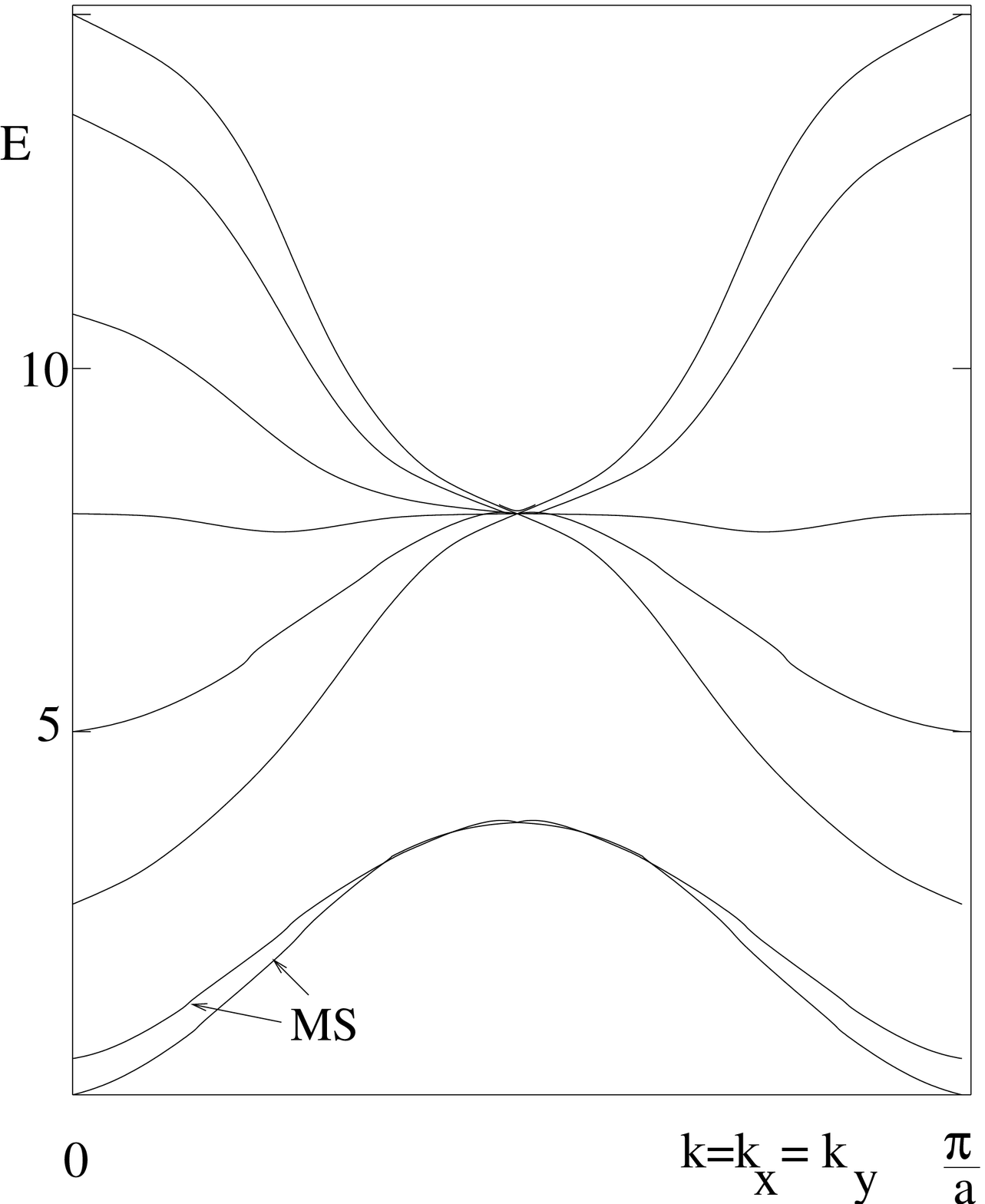}
\caption{\label{ffig16_6}
Left: Magnon spectrum $E=\hbar\omega$ of a ferromagnetic film with a simple cubic lattice versus
$k\equiv k_x=k_y$ for $N_T=8$ and $D/J=0.01$.  No surface mode is observed for this case. Right: Magnon spectrum $E=\hbar\omega$ of a ferromagnetic film with a body-centered cubic lattice versus
$k\equiv k_x=k_y$ for $N_T=8$ and $D/J=0.01$. The branches of surface modes are indicated by  MS.
}
\end{figure}

Let us show in Fig. \ref{ffig16_6} two examples of SW spectrum, one without surface modes as in a simple cubic film and the other with surface localized modes as in body-centered cubic ferromagnetic case.

Note that a surface mode has a damping SW amplitude when going from the surface to the interior. The SW amplitudes for each mode are in fact their eigenvectors calculated from Eq. (\ref{gnn}). It is very important to note that acoustic surface localized spin waves lie below the bulk frequencies so that these low-lying energies will give larger integrands to the integral on the right-hand side of Eq. (\ref{lm2}), making $<S_n^z>$ to be smaller. The same effect explains the diminution of $T_c$ in thin films whenever low-lying surface spin waves exist in the spectrum.

\begin{figure}[h!]
\centering
\includegraphics[width=10 cm]{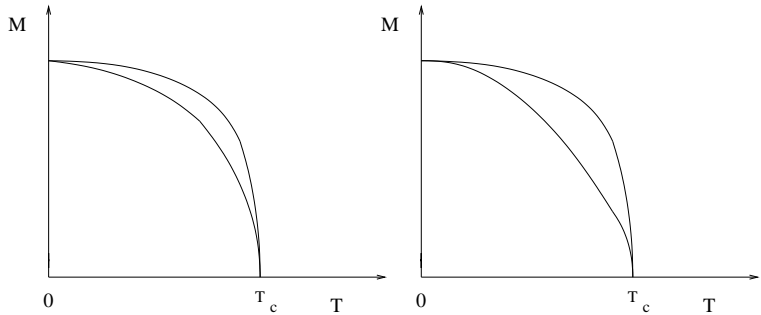}
\caption{\label{ffig16_8}
Ferromagnetic films of simple cubic lattice (left) and body-centered cubic lattice (right): magnetizations of the surface layer
 (lower curve) and  the second layer (upper curve), with $N_T=4$, $D=0.01J$, $J=1$.
}
\end{figure}

Figure \ref{ffig16_8}  shows the results of the layer magnetizations for the first two layers in the films considered above with  $N_T=4$.

Calculations for antiferromagnetic thin films and other cases with non-collinear spin configurations can be performed using generalized Green's functions \cite{Diep1979,DiepTF91,Quartu1998} with the general Hamiltonian defined for
two spins $\mathbf S_i$ and $\mathbf S_j$ forming an angle $\cos \theta_{ij}$:  one can express the Hamiltonian in the local coordinates as follows \cite{Diep2015}

\begin{eqnarray}
\mathcal H &=& - \sum_{<i,j>}
J_{i,j}\Bigg\{\frac{1}{4}\left(\cos\theta_{ij} -1\right)
\left(S^+_iS^+_j +S^-_iS^-_j\right)\nonumber\\
&+& \frac{1}{4}\left(\cos\theta_{ij} +1\right) \left(S^+_iS^-_j
+S^-_iS^+_j\right)\nonumber\\
&+&\frac{1}{2}\sin\theta_{ij}\left(S^+_i +S^-_i\right)S^z_j
-\frac{1}{2}\sin\theta_{ij}S^z_i\left(S^+_j
+S^-_j\right)\nonumber\\
&+&\cos\theta_{ij}S^z_iS^z_j\Bigg\}- \sum_{<i,j>}I_{i,j}S^z_iS^z_j\cos \theta_{ij}
\label{eq:HGH2}
\end{eqnarray}
The last term is an anisotropy added to facilitate a numerical convergence for  ultra thin films at long-wave lengths since it is known that in 2D there is no ordering for isotropic Heisenberg spins at finite temperatures \cite{Mermin}.

The determination of the angles in the ground state can be done either by minimizing the interaction energy with respect to interaction parameters \cite{NgoSurface,NgoSurface2}.  Using their values, one can follow the different steps presented above for the collinear magnetic films, one then obtains a matrix which can be numerically diagonalized  to get the spin-wave spectrum which is used in turn to calculate physical properties in the same manner as for the collinear case presented above.

\section{Frustrated thin films: surface phase transition}\label{surfaces}
Having given the background in the previous section, we can give some results here. The reader is referred to the original papers for lengthy technical details. Our aim here is to discuss physical effects due to the conditions of the surface.

As said earlier, the combination of the frustration and the surface effect gives rise to drastic effects as seen in the examples shown in the following.

\subsection{Frustrated surfaces}

We consider an example in this section: a ferromagnetic film with frustrated surfaces \cite{Ngo2007}. We study,  by  the analytical
Green's function method and extensive Monte Carlo simulations, effects of frustrated surfaces  on the
properties of thin films made of stacked triangular layers of
atoms bearing Heisenberg spins with an Ising-like interaction
anisotropy. We suppose that the in-plane surface interaction $J_s$
 can be antiferromagnetic  or ferromagnetic while all other interactions are
ferromagnetic. We show that the ground-state spin configuration is
non linear when $J_s$ is lower than a critical value $J_s^c$. The
film surfaces are then frustrated. In the frustrated case, there
are two phase transitions related to disordering of surface and
interior layers. There is a good agreement between Monte Carlo and Green's function
results.
\subsection{Model}
We consider a thin film made up by stacking $N_z$
planes of triangular lattice of $L\times L$ lattice sites.

The Hamiltonian is given by
\begin{equation}
\mathcal H=-\sum_{\left<i,j\right>}J_{i,j}\mathbf S_i\cdot\mathbf
S_j -\sum_{<i,j>} I_{i,j}S_i^z S_j^z  \label{eqn:hamil1}
\end{equation}
where $\mathbf S_i$ is the Heisenberg spin at the lattice site
$i$, $\sum_{\left<i,j\right>}$ indicates the sum over the nearest neighbor spin
pairs  $\mathbf S_i$ and $\mathbf S_j$.  The last term, which will
be taken to be very small,  is needed to ensure that there is a phase transition at a finite
temperature for the film with a
finite thickness  when all exchange interactions $J_{i,j}$
are ferromagnetic. Otherwise, it is known that a
strictly two-dimensional system with an isotropic non-Ising spin
model (XY or Heisenberg model) does not have a long-range ordering
at finite temperatures \cite{Mermin}.

We suppose that the interaction between two nearest neighbors on the  surface is equal to $J_s$, and all other interactions are ferromagnetic and  equal to $J=1$ for
simplicity. The two surfaces of the film are frustrated if $J_s$
is antiferromagnetic ($J_s<0$), due to the triangular lattice structure.
\subsection{Ground state}
In this paragraph, we suppose that the spins are classical.  The
classical ground state can be easily determined as shown
below.  Note that for antiferromagnetic systems, even for bulk
materials, the quantum ground state though not far from the classical one, cannot be exactly determined because of the quantum fluctuations \cite{DiepTM}.

 For $J_s>0$
(ferromagnetic interaction), the magnetic ground state is ferromagnetic.
However, when $J_s$ is negative there is a competition between the non collinear surface ordering and the ferromagnetic ordering of the spins of the beneath layer.

\begin{figure}[h!]
\centering
\includegraphics[width= 3 in]{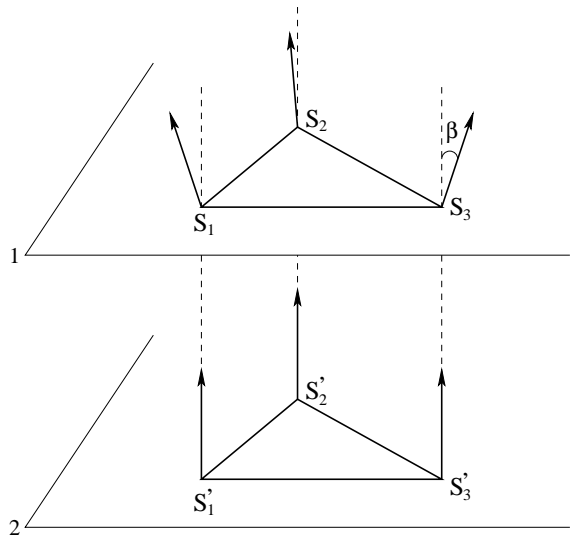}
\caption{Non
collinear surface spin configuration. Angles between spins on
layer $1$ are all equal (noted by $\alpha$), while angles between
vertical spins are $\beta$.} \label{fig:gsstruct}
\end{figure}

We first determine the ground state configuration for $I=I_s=0.1$ by using
the steepest descent method : starting from a random spin
configuration, we calculate the magnetic local field at each site
and align the spin of the site in its local field. In doing so for
all spins and repeating until the convergence is reached, we obtain
in general the ground state configuration, without metastable states in the
present model. The result shows that when $J_{s}$ is smaller than
a critical value $J_{s}^c$ the magnetic ground state is obtained from the
planar $120^\circ$ spin structure in the $XY$
plane, by pulling them out of the  $xy$ plane by an angle
$\beta$. The three spins on a triangle on the surface form thus an
``umbrella" with an angle $\alpha$ between them and an angle
$\beta$ between a surface spin and its beneath neighbor (see Fig.
\ref{fig:gsstruct}). This non planar structure is due to the
interaction of the spins on the beneath layer, just like an
external applied field in the $z$ direction. Of course, when $|J_s|$
is smaller than $|J_s^c|$ one has the collinear ferromagnetic ground state as
expected: the frustration is not strong enough to resist the
ferromagnetic interaction from the beneath layer.

\begin{figure}[h!]
\centering
\includegraphics[width=2.7 in]{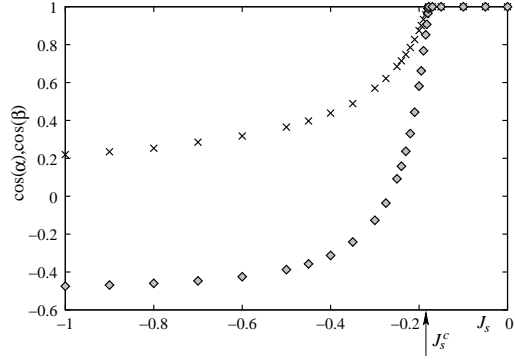}
\caption{$\cos
(\alpha)$ (diamonds) and $\cos (\beta)$ (crosses) as functions of
$J_s$. Critical value of $J_s^c$ is shown by the arrow.}
\label{fig:gscos}
\end{figure}

We show in Fig. \ref{fig:gscos} $\cos(\alpha)$ and $\cos(\beta)$
as functions of $J_s$. The critical value $J_s^c$ is found between
-0.18 and -0.19.  This value can be calculated analytically by
assuming the ``umbrella structure". For ground state analysis, it suffices to
consider just a cell shown in Fig. \ref{fig:gsstruct}. This is
justified by the numerical determination discussed above.
Furthermore, we consider as a single solution all configurations
obtained from each other by any global spin rotation.

Let us consider the full Hamiltonian (\ref{eqn:hamil1}).  For
simplicity, the interaction inside the surface layer is set equal
$J_s$ $(-1 \leq J_s \leq 1)$ and all others are set equal to
$J>0$. Also, we suppose that $I_{i,j}=I_s$ for spins on the
surfaces with the same sign as $J_s$ and all other $I_{i,j}$ are
equal to $I>0$ for the inside spins including interaction between
a surface spin and a nn spin on the beneath layer.

The spins are numbered as in Fig. \ref{fig:gsstruct}: $S_1$, $S_2$
and $S_3$ are the spins in the surface layer (first layer),
$S'_1$, $S'_2$ and $S'_3$ are the spins in the internal layer
(second layer).  The Hamiltonian for the cell is written as
\begin{eqnarray}
H_p &=& -6\left[ J_s\left( \mathbf S_1\cdot \mathbf S_2 +\mathbf
S_2\cdot\mathbf S_3 + \mathbf S_3\cdot\mathbf S_1
\right)\right.\nonumber\\
&&+I_s\left( S^z_1S^z_2 + S^z_2S^z_3 + S^z_3S^z_1\right)\nonumber\\
&+&J\left(\mathbf S'_1\cdot \mathbf S'_2 +\mathbf
S'_2\cdot\mathbf S'_3 +\mathbf S'_3\cdot\mathbf S'_1\right)\nonumber\\
&&+I\left.\left( S'^z_1S'^z_2 + S'^z_2S'^z_3 + S'^z_3S'^z_1\right)\right] \nonumber \\
&-&2J\left( \mathbf S_1\cdot \mathbf S'_1 +\mathbf S_2\cdot\mathbf
S'_2 +\mathbf S_3\cdot\mathbf S'_3\right)\nonumber\\
&&-2I\left( S^z_1S'^z_1 + S'^z_2S'^z_2 + S^z_3S'^z_3\right),
\label{eqn:Hamilplaq}
\end{eqnarray}
Let us decompose each spin into two components: an $xy$ component,
which is a vector, and a $z$ component $\mathbf S_i=(\mathbf
S_i^{\parallel}, S_i^z)$. Only surface spins have $xy$ vector
components.  The angle between these $xy$ components of  nearest neighbor
surface spins is $\gamma_{i,j}$ which is chosen by ($\gamma_{i,j}$
is in fact the projection of $\alpha $ defined above on the $xy$
plane)
\begin{equation}
\gamma_{1,2}=0,\ \gamma_{2,3}=\frac{2\pi}{3},\
\gamma_{3,1}=\frac{4\pi}{3}. \label{eqn:HSAngAlpha}
\end{equation}

The angles $\beta_i$ and $\beta'_i$ of the spin $\mathbf S_i$ and
$\mathbf S'_i$ with the $z$ axis are by symmetry
$$
\left\{%
\begin{array}{c}
\beta_1=\beta_2=\beta_3=\beta,\\
\beta'_1=\beta'_2=\beta'_3=0,\\
\end{array}
\right.
$$

The total energy of the cell (\ref{eqn:Hamilplaq}), with $S_i =
S'_i = \frac{1}{2}$, can be rewritten as
\begin{eqnarray}
H_p&=&-\frac{9(J+I)}{2} -\frac{3(J+I)}{2}\cos\beta
-\frac{9(J_s+I_s)}{2}\cos^2\beta
\nonumber\\
&+&\frac{9J_s}{4}\sin^2\beta. \label{eqn:totEplaq}
\end{eqnarray}
By a variational method,  the minimum of the cell energy
corresponds to
\begin{equation}
\frac{\partial H_p}{\partial\beta} =\left( \frac{27}{2}J_s+
9I_s\right)\cos\beta\sin\beta +\frac{3}{2}(J+I)\sin\beta \ = \ 0
\label{eqn:DerivE}
\end{equation}
We have
\begin{equation}
\cos\beta = -\frac{J+I}{9J_s+6I_s}. \label{eqn:GSsolu}
\end{equation}

For  given values of $I_s$ and $I$, we see that the solution
(\ref{eqn:GSsolu}) exists for $J_s \leq J_s^c$ where the critical
value $J_s^c$ is determined by $-1\leq \cos\beta \leq 1$. For
$I=-I_s=0.1$, one obtains $J_s^c \approx -0.1889 J $
in excellent agreement with the numerical result.

The classical ground state determined here will be used as input ground state
configuration for quantum spins in the calculation by the Green's method.

\subsection{Results from the Green's function method}\label{GreenATL}

Let us consider the quantum spin case.   The details of
the method in the case of non collinear spin configuration have
been given in Ref. \cite{Ngo2007}. We just show the results on the surface phase transition and compare with the Monte Carlo results performed on the equivalent classical model.

\subsubsection{Phase transition and phase diagram of the quantum case}

\begin{figure}[h!]
\centering
\includegraphics[width=2.8 in]{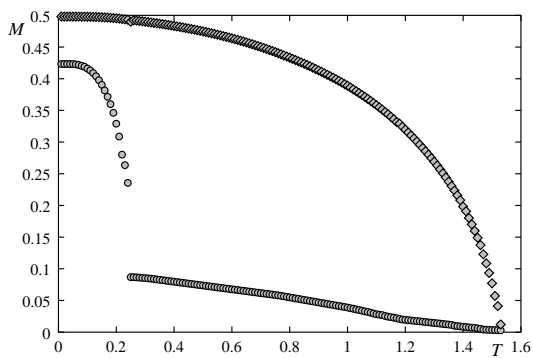}
\caption{First
two layer-magnetizations obtained by the Green's function technique
vs. $T$ for $J_{s} = -0.5$ with $I=-I_s=0.1$. The surface-layer
magnetization (lower curve) is much smaller than the second-layer
one. See text for comments.} \label{fig:HGn05Ms}
\end{figure}

We first show an example where $J_{s} = -0.5$ in Fig.
\ref{fig:HGn05Ms}. As seen, the surface-layer  magnetization is
much smaller than the second-layer one. In addition there is a
strong spin contraction at $T=0$ for the surface layer. This is
due to the antiferromagnetic nature of the in-plane surface
interaction $J_s$ \cite{DiepTM}.  One sees that the surface becomes disordered
at a temperature $T_1\simeq 0.2557$ while the second layer remains
ordered up to $T_2\simeq 1.522$.   Therefore, the system is
partially disordered for temperatures between $T_1$ and $T_2$.
 This result is very interesting because it confirms again the
existence of the partial disorder in quantum spin systems observed
earlier in the bulk \cite{Rocco,santa2}.  Note that between $T_1$ and $T_2$,
the ordering of the second layer acts as an external field on the
first layer, inducing therefore a small value of its
magnetization.  A further evidence of the existence of the surface
transition will be provided with the surface susceptibility in the
Monte Carlo results shown below.

\begin{figure}[h!]
\centering
\includegraphics[width=2.7 in]{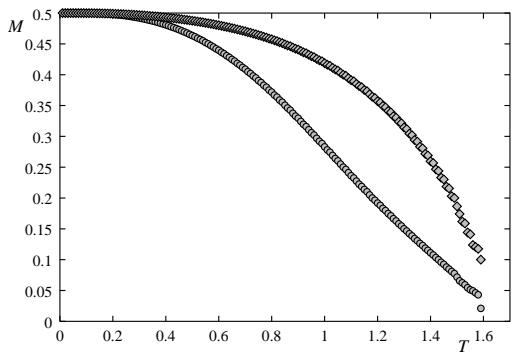}
\caption{First
two layer-magnetizations obtained by the Green's function technique
vs. $T$ for $J_{s} = 0.5$ with $I=I_s=0.1$.} \label{fig:HGp05Ms}
\end{figure}

Figure \ref{fig:HGp05Ms} shows the non frustrated case where
$J_s=0.5$, with $I=I_s=0.1$.  As seen, the first-layer
magnetization is smaller than the second-layer one. There is only
one transition temperature. Note the difficulty for numerical
convergency when the magnetizations come close to zero.

\begin{figure}[h!]
\centering
\includegraphics[width=2.7 in]{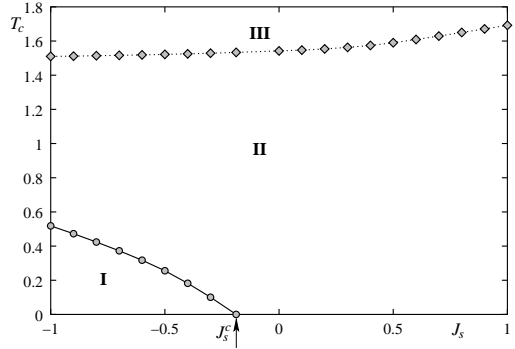}
\caption{Phase
diagram in the space ($J_{s},T$) for the quantum Heisenberg model
with $N_z=4$, $I=|I_s|=0.1$. See text for the description of
phases I to III.} \label{fig:HGDG}
\end{figure}

We show in Fig. \ref{fig:HGDG} the phase diagram in the space
$(J_s,T)$.  Phase I denotes the ordered phase with surface non
collinear spin configuration, phase II indicates the collinear
ordered state, and phase III is the paramagnetic phase. Note that
the surface transition does not exist for $J_s \geq J_s^c$.

\subsubsection{Monte Carlo results}

The Green's function method can go up to
$T_c$ but due to the decoupling scheme, it cannot give a correct
critical behavior at $T_c$.  An alternative method for high temperatures
is to consider the counterpart classical spins and to use Monte Carlo
simulations to obtain the phase
diagram for comparison.  This is somewhat justified because the quantum nature of spins
is no more important at high $T$.

For Monte Carlo simulations (see methods in Refs. \cite{Metropolis,Binder,Ferrenberg,Ferrenberg2,DiepSP}),  we use the same Hamiltonian (\ref{eqn:hamil1}) but the spins are the classical Heisenberg model of magnitude $S=1$.
The film sizes are $L\times L \times N_z$ where
$N_z=4$ is the number of layers (film thickness) taken as in the
quantum case presented above. We use here $L=24, 36, 48, 60$ to
study finite-size effects. Periodic boundary
conditions are used in the $xy$ planes. The equilibrating time is
about $10^6$ Monte Carlo steps per spin and the averaging time is $2\times
10^6$ Monte Carlo steps per spin. $J=1$ is taken as unit of energy in the
following.

\begin{figure}[h!]
\centering
\includegraphics[width=2.7 in]{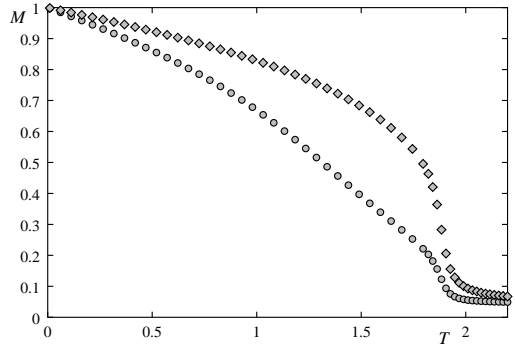}
\caption{Magnetizations of layer 1 (circles) and layer 2
(diamonds) versus temperature $T$ in unit of $J/k_B$ for $J_s=0.5$
with $I=I_s=0.1$, $L=36$.} \label{fig:HSp05Ms}
\end{figure}

Figure \ref{fig:HSp05Ms} shows the layer magnetizations of
the first two layers as a function of $T$ , in the case $J_s=0.5$ (no frustration)
with $N_z=4$ (the third and fourth layers are symmetric). In this case, there is
clearly no surface transition just as in the quantum case.

\begin{figure}[h!]
\centering
\includegraphics[width=2.7 in]{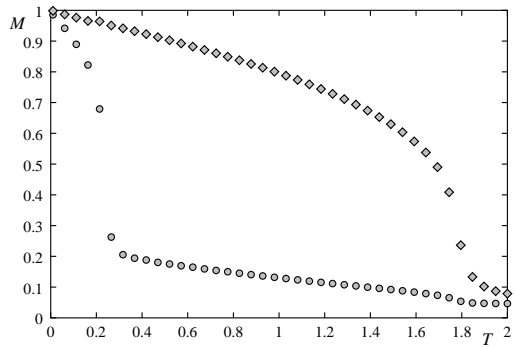}
\caption{Magnetizations of layer 1 (circles) and layer 2
(diamonds) versus temperature $T$ in unit of $J/k_B$ for
$J_s=-0.5$ with $I=-I_s=0.1$, $L=36$.} \label{fig:HSn05Ms}
\end{figure}

Figure \ref{fig:HSn05Ms} shows a frustrated case where
$J_s=-0.5$.  The surface layer in this case becomes disordered at
a temperature much lower than that for the second layer.  Note
that the surface magnetization is slightly smaller than 1 at $T=0$ (not seen with the scale of the figure).
This is  because the surface spins make an angle with the $z$ axis
so their $z$ component is less than 1 in the ground state.

\begin{figure}[h!]
\centering
\includegraphics[width=2.7 in]{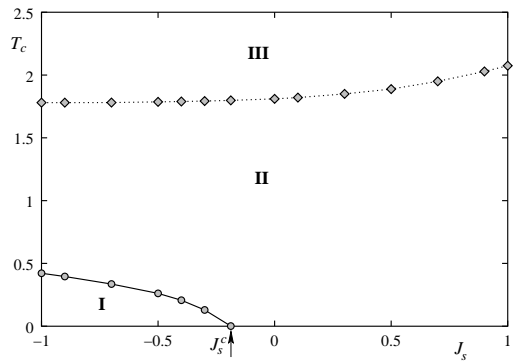}
\caption{Phase
diagram in the space ($J_s,T$) for the classical Heisenberg model
with $N_z=4$, $I=|I_s|=0.1$. Phases I to III have the same
meanings as those in Fig. \ref{fig:HGDG} .} \label{fig:HSDG}
\end{figure}

The phase diagram is shown in Fig.
\ref{fig:HSDG} in the space
$(J_s,T)$.  This phase diagram resembles
remarkably to that obtained for the quantum counterpart model
shown in Fig. \ref{fig:HGDG}.

\subsection{Frustrated thin films}
 We have also studied frustration effects in antiferromagnetic fcc Heisenberg films \cite{NgoSurface2}. In this case, the whole film is frustrated due to the geometry of the lattice.

 Let us consider a film of fcc lattice structure with (001)
surfaces. To avoid the absence of long-range order of isotropic
non Ising spin model at finite $T$ when the film
thickness is very small, i.e. quasi 2D
system \cite{Mermin}, we add in the Hamiltonian an Ising-like
uniaxial anisotropy term. The Hamiltonian is given by
\begin{equation}
\mathcal H=-\sum_{\left<i,j\right>}J_{i,j}\mathbf S_i\cdot\mathbf
S_j -\sum_{i} D_{i}(S^z_i)^2  \label{eqn:hamil2}
\end{equation}
where $\mathbf S_i$ is the Heisenberg spin at the lattice site
$i$, $\sum_{\left<i,j\right>}$ indicates the sum over the nn spin
pairs  $\mathbf S_i$ and $\mathbf S_j$.

In the following, the interaction between two nn surface spins is
denoted by $J_s$, while all other interactions are supposed to be
antiferromagnetic and all equal to $J=-1$ for simplicity.

\begin{figure}[h!]
\centering
\includegraphics[width=3.2 in]{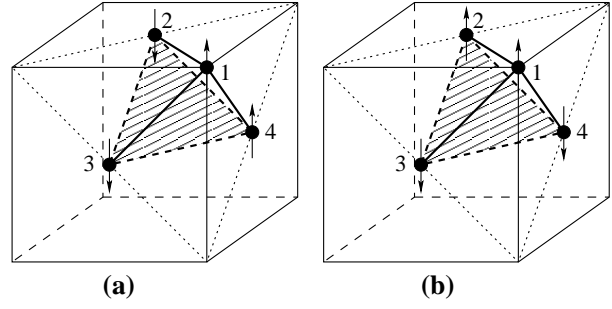}
\caption{The
ground state spin configuration of the fcc cell at the film
surface: a) ordering of type I for $J_s<-0.5$; b) ordering of type
II for $J_s>-0.5$.} \label{fig:gsstruct1}
\end{figure}

The ground state is shown (see demonstration in Ref. \cite{NgoSurface2}) to depend on the surface in-plane interaction $J_s$ with a critical value $J^c_s =-0.5$ at which ordering of type I coexists with ordering of type II (see Fig. \ref{fig:gsstruct1}).

For $J_s < J^c_s$, the
spins in each $yz$ plane are parallel while spins in adjacent $yz$
planes are antiparallel (Fig. \ref{fig:gsstruct1}a). This ordering
will be called hereafter "ordering of type I": in the $x$ direction the
ferromagnetic planes are antiferromagnetically coupled as shown in this
figure. Of course, there is a degenerate configuration where the ferromagnetic planes are antiferromagnetically ordered in the $y$ direction.  Note that the surface layer
has an antiferromagnetic ordering for both configurations.  The degeneracy of  type
I is therefore 4 including the reversal of all spins.

For $J_s
> J^c_s$, the spins in each $xy$ plane is
ferromagnetic. The adjacent $xy$ planes have an antiferromagnetic ordering in the $z$
direction perpendicular to the film surface.   This
will be called hereafter "ordering of type II".  Note that the
surface layer is then ferromagnetic (Fig. \ref{fig:gsstruct1}b). The
degeneracy of  type II is 2 due to the reversal of all spins.

\begin{figure}[h!]
\centering
\includegraphics[width=2.8 in]{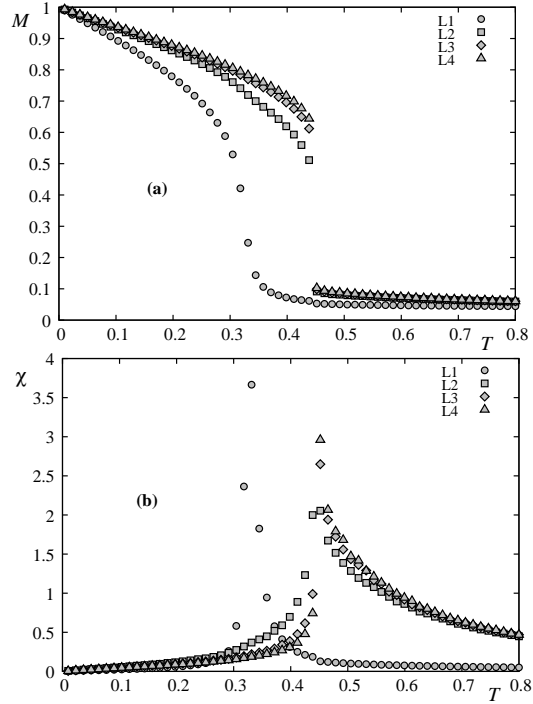}
\caption{Magnetizations and susceptibilities of first two cells vs
temperature for $J_s = -0.5$ with $D = 0.1$. $L_j$
denotes the sublattice magnetization of layer $j$. The
susceptibility of sublattice 1 of the first cell is divided by a
factor $5$ for presentation convenience.} \label{fig:N24D01J05}
\end{figure}

\begin{figure}[h!]
\centering
\includegraphics[width=2.8 in]{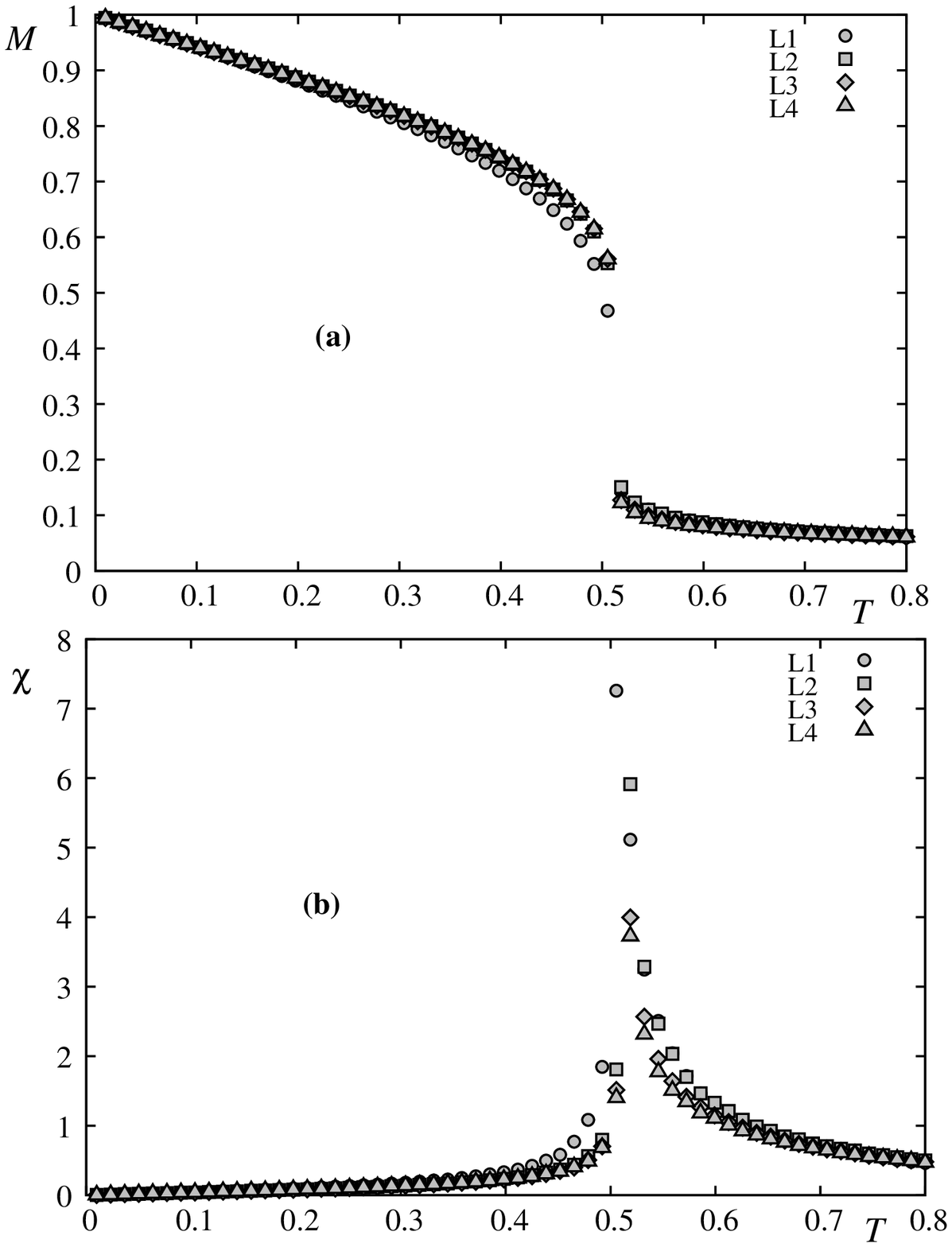}
\caption{Magnetizations and susceptibilities of sublattices 1 and
3 of first two cells vs temperature for $J_s = -0.8$ with $D = 0.1$. $L_j$ denotes the sublattice magnetization of layer
$j$.} \label{fig:N24D01J08}
\end{figure}

\begin{figure}[h!]
\centering
\includegraphics[width=2.8 in]{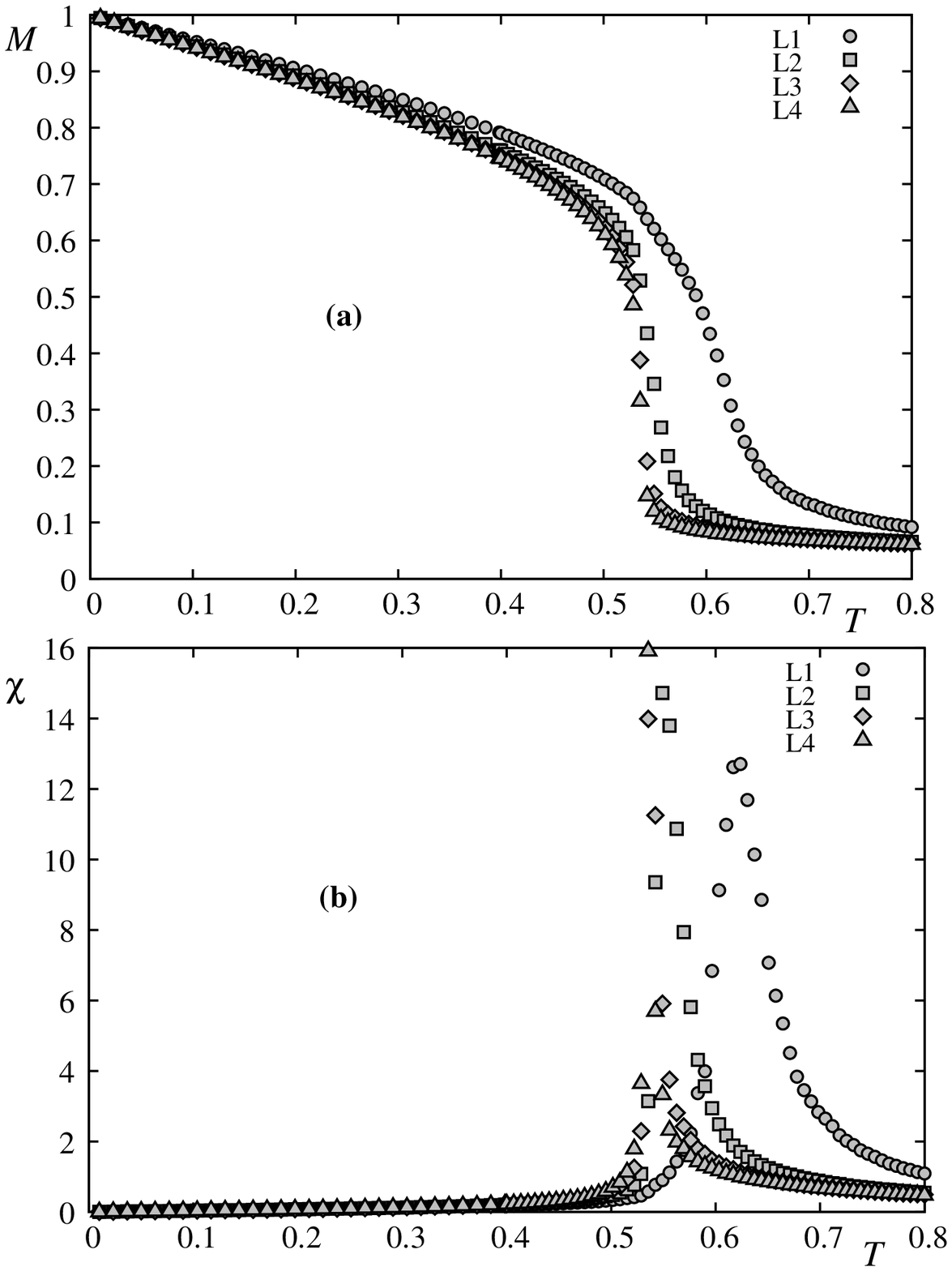}
\caption{Magnetizations and susceptibilities of sublattices 1 and
3 first two cells vs temperature for $J_s = -1.0$ with
$D = 0.1$. $L_j$ denotes the sublattice magnetization of layer
$j$.} \label{fig:N24D01J10}
\end{figure}

Monte Carlo simulations have been used to study the phase transition in this frustrated film.  We just show below three typical cases, at and far from $ J^c_s$. Figure \ref{fig:N24D01J05} shows the sublattice layer magnetizations at $J_s^c=-0.5$ where one sees that the surface layer undergoes a transition at a temperature lower than the interior ones.  Far from this value there is a single phase transition as seen in Fig. \ref{fig:N24D01J08}.  However, when $J_s$ is negatively stronger, we have a hard surface, namely the surface undergoes a phase transition at a $T$ higher than the interior layer, as seen in Fig. \ref{fig:N24D01J10}


\begin{figure}[h!]
\centering
\includegraphics[width=2.8 in]{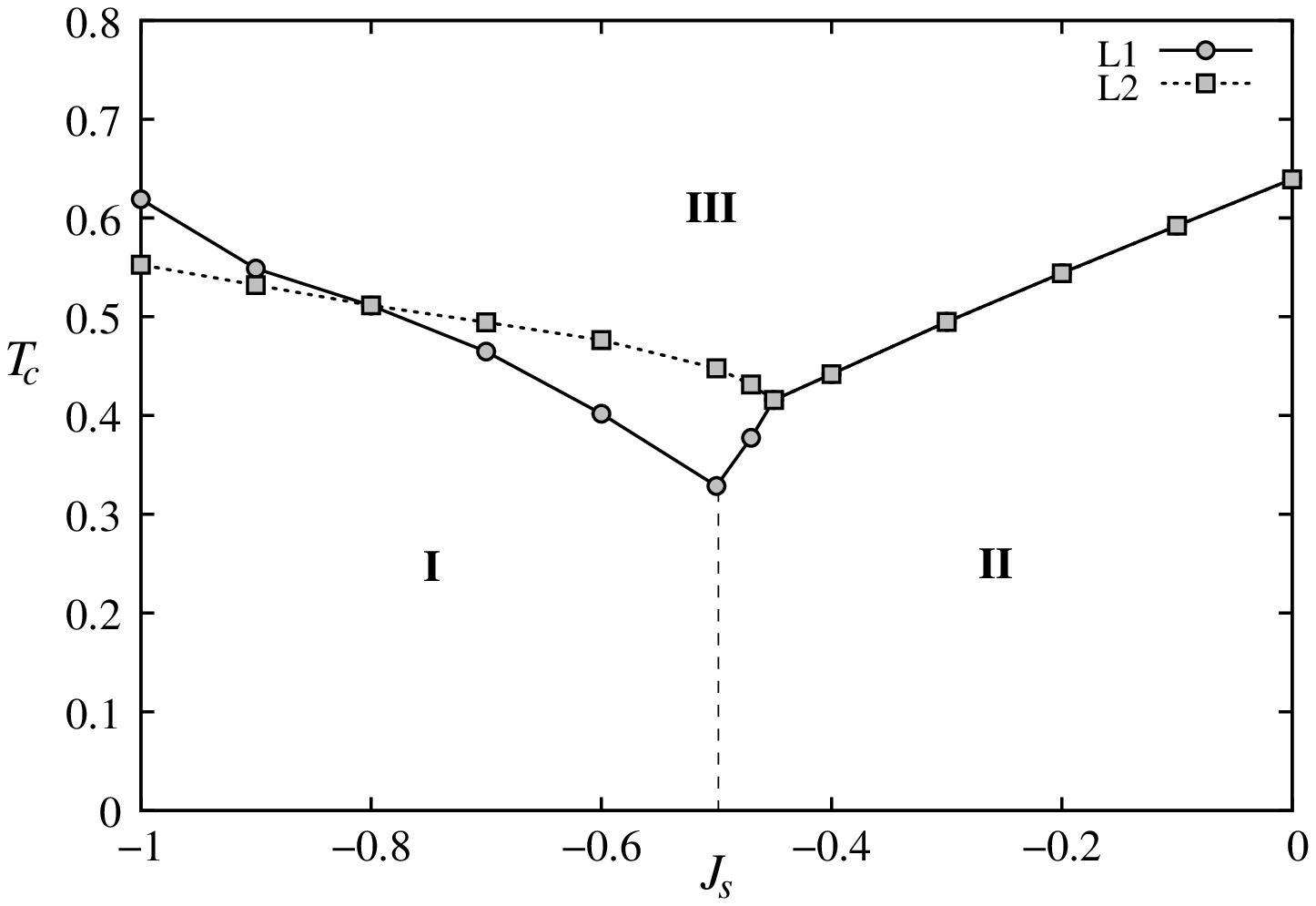}
\caption{Phase diagram in the space $(J_s,T_c)$ with $D = 0.1$.
$L_j$ denotes data points for the maximum of the sublattice
magnetization of layer $j$.  I and II denote ordering of type I
and II defined in in Fig. \ref{fig:gsstruct1}.  III is paramagnetic
phase.  The discontinued vertical line is a first-order line.
Errors are smaller than symbol sizes. See text for comments. }
\label{fig:PDJ24D01}
\end{figure}

The phase diagram is shown in Fig. \ref{fig:PDJ24D01}

  Note that near the phase boundary $J_s^c$ ($-0.5\leq J_s \leq -0.43$) a reentrant phase is found between phases I and II (not seen with the figure scale).  As said in the 2D exactly solved models above,  one has to be careful to examine a very small region near the phase boundary where unexpected phenomena can occur. This is the case here.

  The nature of the phase transition is also studied by a histogram technique \cite{Ferrenberg,Ferrenberg2}.
  Critical exponents are found to have values between 2D and 3D universality classes.
  The reader is referred to Ref. \cite{NgoSurface2} for details.
We will return to this point in section \ref{critical} below.

\subsection{Helimagnetic films}
Bulk helimagnets have been studied a long time ago \cite{Harada,Diep1989a,Diep1989b}.   A simple helimagnetic order resulting from the competition between the nn and nnn interactions is  shown in section \ref{noncollinear}.  Helimagnetic films are seen therefore as frustrated films.

We have recently used the Green's function method and Monte Carlo simulations to study helimagnetic films in zero field \cite{Diep2015,Diep2016} and in a perpendicular field \cite{DiepField2017}.  We summarize here some results and emphasize their importance.

Consider the following helimagnetic Hamiltonian

\begin{equation}
\mathcal H=-\sum_{\left<i,j\right>}J_{i,j}\mathbf S_i\cdot\mathbf S_j -\sum_i \mathbf H\cdot \mathbf S_i
 \label{eqn:hamil3}
\end{equation}
where $J_{i,j}$ is the interaction between two spins $\mathbf S_i$ and $\mathbf S_j$ occupying the lattice sites $i$ and $j$ and $\mathbf H$ denotes an external magnetic field applied along the $c$ axis.  To generate helical angles in the $c$ direction, we suppose an antiferromagnetic interaction $J_2$ between nnn in the $c$ direction in addition to the ferromagnetic interaction $J_1$ between nn in all directions. For simplicity, we suppose that $J_1$ is the same everywhere.
For this section we shall suppose $J_2$ is the same everywhere for the presentation clarity.  Note that in the bulk in zero field, the helical angle along the $c$ axis is given by $\cos \alpha=-\frac{J_1}{4J_2}$ for a simple cubic lattice \cite{DiepTM} with $|J_2|>0.25 J_1$. Below this value, the ferromagnetic ordering is stable.

\begin{figure}[h!]
\centering
\includegraphics[width=7cm,angle=0]{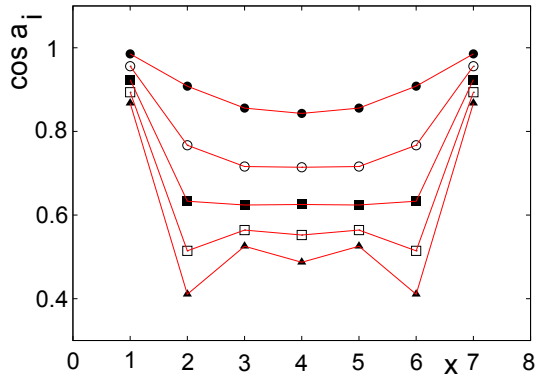}  
\caption{Cosinus of $\alpha_1=\theta_1-\theta_2$, ..., $\alpha_7=\theta_7-\theta_8$ across the film
for $J_2/J_1=-1.2,-1.4,-1.6,-1.8, -2$ (from top) with thickness $N_z=8$: $a_i$ stands for $\theta_i-\theta_{i+1}$ and $x$ indicates the film layer $i$ where the angle $a_i$ with the layer $(i+1)$ is shown.  A strong rearrangement of spins near the surface is observed. }\label{GSA}
\end{figure}

In zero field the helical angle has been shown to be strongly modified near the surface as shown in Fig. \ref{GSA}

\begin{figure}[htb]
\centering
\includegraphics[width=7cm,angle=0]{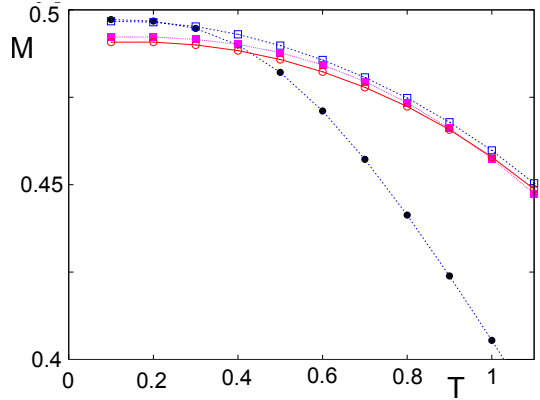}  
\caption{(Color online) Layer magnetizations as functions of $T$ for $J_2/J_1=-1.4$ with $d=0.1$, thickness $N_z=8$. Note the crossover at low $T$. Black circles, blue void squares, magenta squares and red void circles are for first, second, third and fourth layers, respectively.  See text.}\label{magnet14}
\end{figure}

Some results from the laborious Green's function are shown in Fig. \ref{magnet14}. Note the crossover of the layer magnetizations at low $T$. This is due to quantum fluctuations which are different for each layer, depending on the antiferromagnetic interaction strength (namely the so-called zero-point spin contractions, see Ref. \cite{DiepTM}). Without such a theoretical insight, it would be difficult to analyze experimental data when this happens.

In an applied field \cite{DiepField2017}, we have observed a new phenomenon, namely a partial phase transition in the helimagnetic film: contrary to what has been shown above (surface phase transition below or above the bulk one), here we have each single interior layer undergoes a separate transition.
Synthetically, we can summarize that under the applied magnetic field,  the spins in the GS make different angles between them across the film. When the temperature increases, the layers with large $xy$ spin-components undergo a  phase transition where the transverse (in-plane) $xy$ ordering is destroyed.  This "transverse" transition is possible because the $xy$ spin-components do not depend on the field.  Other layers with small $xy$ spin-components, namely large $z$ components, do not make a transition because the ordering in $S^z$ is maintained by the applied field.  The transition of a number of layers with large $xy$ spin-components, not all layers, is a new phenomenon discovered here with our present model.
Experiments have been performed on materials with  helical structures often more complicated than the model considered in this paper. However, the clear physical pictures given in our present analysis are believed to be useful in the search for the interpretation of experimental data.

\section{Criticality of thin films}\label{critical}

One of the important fundamental questions in surface physics is the criticality of the phase transition in thin films.

To answer this question, we studied  the critical behavior of magnetic thin films as a function of the film thickness \cite{DiepCritical}.  We used the ferromagnetic Ising
model with the high-resolution multiple histogram Monte Carlo
simulation \cite{Ferrenberg,Ferrenberg2}. We showed that though the 2D behavior remains dominant
at small thicknesses, there is a systematic continuous deviation
of the critical exponents from their 2D values.
 We explain these deviations using the concept of
"effective" exponents suggested by Capehart and Fisher \cite{Fisher} in a
finite-size analysis.  The shift of the critical temperature with
the film thickness obtained here by Monte Carlo simulation is in an
excellent agreement with their prediction.

We summarize here this work.

Let us consider the Ising spin model on a film
made from a ferromagnetic simple cubic lattice. The size of the
film is $L\times L\times N_z$.  We apply the periodic boundary
conditions (PBC) in the  $xy$ planes to simulate an infinite $xy$
dimension. The $z$ direction is limited by the film thickness
$N_z$.   If $N_z=1$ then one has a 2D square lattice.

The Hamiltonian is given by
\begin{equation}
\mathcal H=-\sum_{\left<i,j\right>}J_{i,j}\sigma_i\cdot\sigma_j
\label{eqn:hamil4}
\end{equation}
where $\sigma_i$ is the Ising spin of magnitude 1 occupying the
lattice site $i$, $\sum_{\left<i,j\right>}$ indicates the sum over
the nn spin pairs  $\sigma_i$ and $\sigma_j$.

Using the high-precision multi-histogram Monte Carlo technique \cite{Ferrenberg,Ferrenberg2} we have calculated various critical exponents as functions of the film thickness using the finite-size scaling \cite{Barber} described as follows.
In Monte Carlo simulations, one calculates the averaged order parameter
$\langle M\rangle$ ($M$: magnetization of the system), averaged
total energy $\langle E\rangle$, specific heat $C_v$,
susceptibility $\chi$, first order cumulant of the energy $C_U$,
and $n^{th}$ order cumulant of the order parameter $V_n$ for $n=1$
and 2. These quantities are defined as

\begin{eqnarray}
\langle E\rangle&=&\langle\cal{H}\rangle,\\
C_v&=&\frac{1}{k_BT^2}\left(\langle E^2\rangle-\langle E\rangle^2\right),\\
\chi&=&\frac{1}{k_BT}\left(\langle M^2\rangle-\langle M\rangle^2\right),\\
C_U&=&1-\frac{\langle E^4\rangle}{3\langle E^2\rangle^2},\\
V_n&=&\frac{\partial\ln{M^n}}{\partial(1/k_BT)} =\langle
E\rangle-\frac{\langle M^nE\rangle}{\langle M^n\rangle}.
\end{eqnarray}
Let us discuss the case where all dimensions can go to infinity.
For example, consider a system of size $L^d$ where $d$ is the
space dimension.  For a finite $L$, the pseudo "transition"
temperatures can be identified by the maxima of $C_v$ and $\chi$,
.... These maxima do not in general take place at the same
temperature. Only at infinite $L$ that the pseudo "transition"
temperatures of these respective quantities coincide at the real
transition temperature $T_c(\infty)$. So when we work at the
maxima of $V_n$, $C_v$ and $\chi$, we are in fact working at
temperatures away from $T_c(\infty)$.   This is an important point to bear in mind
for the discussion given below. Let us define the reduced
temperature which measures the "distance" from $T_c(\infty)$ by

\begin{equation}\label{rt}
t=\frac{T-T_c(\infty)}{T_c(\infty)}
\end{equation}
This distance tends to zero when all dimensions go to infinity.
For large values of $L$, the following scaling relations are
expected (see details in Ref. \cite{Bunker}):

\begin{equation}\label{V1}
V_1^{\max}\propto L^{1/\nu}, \hspace{1cm} V_2^{\max}\propto
L^{1/\nu},
\end{equation}
\begin{equation}
C_v^{\max}=C_0+C_1L^{\alpha/\nu}\label{Cv}
\end{equation}
and
\begin{equation}\label{chis}
\chi^{\max}\propto L^{\gamma/\nu}
\end{equation}
at their respective 'transition' temperatures $T_c(L)$, and

\begin{equation}
C_U=C_U[T_c(\infty)]+AL^{-\alpha/\nu},
\end{equation}
\begin{equation}
M_{T_c(\infty)}\propto L^{-\beta/\nu}\label{MB}
\end{equation}
and
\begin{equation}
T_c(L)=T_c(\infty)+C_AL^{-1/\nu}\label{TC},
\end{equation}
where $A$, $C_0$, $C_1$ and $C_A$ are constants. We estimate $\nu$
independently from $V_1^{\max}$ and $V_2^{\max}$. With this value
we calculate $\gamma$ from $\chi^{\max}$ and $\alpha$ from
$C_v^{\max}$.  Note that we can estimate $T_c(\infty)$ using the
last expression. Then, using $T_c(\infty)$, we can calculate
$\beta$ from $M_{T_c(\infty)}$. The Rushbrooke scaling law $\alpha
+2\beta +\gamma=2$ is then in principle verified \cite{Barber}.

The results are shown in Table \ref{tab:criexp} where we observe a systematic deviation of the 2D critical exponents with increasing thickness.

\begin{table*}
  \centering
  \caption{Critical exponents, effective dimension and critical temperature
  at infinite $xy$ limit as obtained in this paper.}\label{tab:criexp}
  \begin{tabular}{| r | c | c | c | c | c | c |}
    \hline
    $N_z$ & $\nu$ & $\gamma$ & $\alpha$ & $\beta$ & $d_{\mathrm{eff}}$ &
$T_c(L=\infty,N_z)$ \\
\hline 1 & $0.9990 \pm 0.0028$ & $1.7520 \pm 0.0062$ & $0.00199
\pm 0.00279$ & $0.1266 \pm 0.0049$ & $2.0000 \pm 0.0028$ &
$2.2699\pm
0.0005$ \\
3 & $0.9922 \pm 0.0019$ & $1.7377 \pm 0.0035$ & $0.00222 \pm
0.00192$ & $0.1452 \pm 0.0040$ & $2.0135 \pm 0.0019$ & $3.6365 \pm
0.0024$ \\
5 & $0.9876 \pm 0.0023$ & $1.7230 \pm 0.0069$ & $0.00222 \pm
0.00234$ & $0.1639 \pm 0.0051$ & $2.0230 \pm 0.0023$ & $4.0234 \pm
0.0028$ \\
7 & $0.9828 \pm 0.0024$ & $1.7042 \pm 0.0087$ & $0.00223 \pm
0.00238$ & $0.1798 \pm 0.0069$ & $2.0328 \pm 0.0024$ & $4.1939 \pm
0.0032$ \\
9 & $0.9780 \pm 0.0016$ & $1.6736 \pm 0.0084$ & $0.00224 \pm
0.00161$ & $0.1904 \pm 0.0071$ & $2.0426 \pm 0.0016$ & $4.2859 \pm
0.0022$ \\
11& $0.9733 \pm 0.0025$ & $1.6354 \pm 0.0083$ & $0.00224 \pm
0.00256$ & $0.1995 \pm 0.0088$ & $2.0526 \pm 0.0026$ & $4.3418 \pm
0.0032$ \\
13& $0.9692 \pm 0.0026$ & $1.6122 \pm 0.0102$ & $0.00226 \pm
0.00268$ & $0.2059 \pm 0.0092$ & $2.0613 \pm 0.0027$ & $4.3792 \pm
0.0034$ \\
    \hline
  \end{tabular}
\end{table*}

\begin{figure}[h!]
\centering
\includegraphics[width=2.8 in]{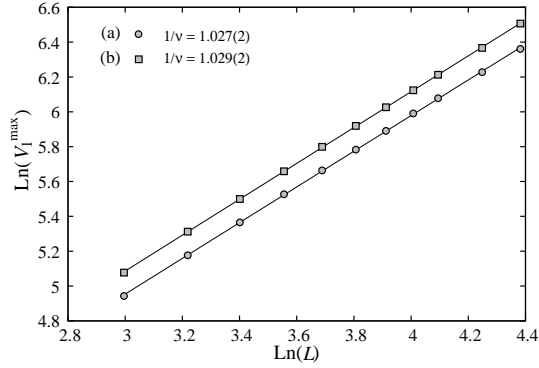}
\caption{Maximum of the first derivative of $\ln M$  versus $ L$
in the $\ln-\ln$ scale for $N_z=11$ (a) without PBC in $z$
direction (b) with PBC in $z$ direction.  The slopes are indicated
on  the figure.  See text for comments. } \label{fig:NUZ11}
\end{figure}

An example of  the calculation of $\nu$ is shown in Fig. \ref{fig:NUZ11} for $N_z=11$ to illustrate the precision of the method: the slope of the "perfect" straight line of our data points gives $1\nu$.

Note that the PBC in the $z$ direction does not change the result if we do not apply the finite-size scaling in that direction \cite{DiepCritical}.

We have also shown that by decreasing the film thickness, a first-order transition in a frustrated fcc Ising thin film can become a second-order transition \cite{DiepCrossover}.

\section{Skyrmions in thin films and superlattices}\label{skyrmions}

Skyrmions are topological excitations in a spin system. They result from the competition between different interactions in an applied magnetic field.

Skyrmions have been discovered by Skyrme \cite{Skyrme} in the context of nuclear physics. Skyrmions have been shown to exist in condensed matter \cite{Bogdanov2003,Leonov,Ackerman,Ezawa,Yu2,Yu1,Muhlbauer,Bauer}.

We  consider in this section the case of a sheet of square lattice of size $N\times N$,  occupied by Heisenberg spins interacting via a nn ferromagnetic exchange interaction $J$ and a nn Dzyaloshinskii-Moriya (DM) interaction \cite{Dzyaloshinskii,Moriya}.
The Hamiltonian is given by
\begin{eqnarray}
\mathcal{H}&=&-J \sum_{\langle ij \rangle} \mathbf {S_i} \cdot \mathbf {S_j} +D \hat z\cdot \sum_i \mathbf {S_i} \wedge (\mathbf{S}_{i+x}  +\mathbf S_{i+y} )\nonumber\\
&&-H \sum_i S_i^z
\end{eqnarray}
where the $ D$ vector of the DM interaction is chosen along the $\hat z$ direction perpendicular to the plane.

In zero field we have studied the spin waves and layer magnetizations at $T=0$ and at finite $T$ \cite{SahbiS}.  The results show that the DM interaction strongly affects the first mode of the spin-wave spectrum.  Skyrmions appear only when an external field is applied perpendicular to the film, as seen in the following.

\begin{figure}[h!]
\centering
\includegraphics[width=3 in]{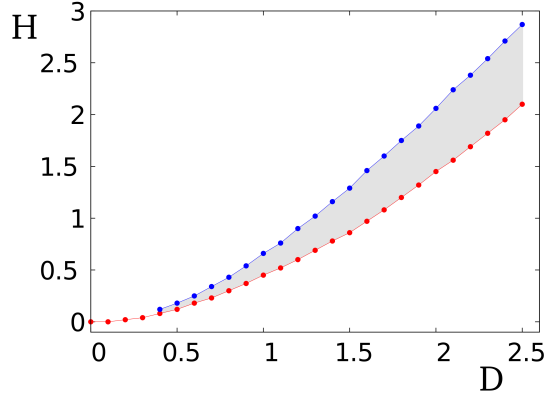}
\caption{Phase diagram in the $(D,H)$ plane for size $N=100$.\label{PD}}
\end{figure}

With $H\neq 0$, we minimize numerically the above Hamiltonian for a given pair ($H,D$), taking $J=1$ as unit, we obtain the GS configuration of the system. The phase diagram is shown in Fig. \ref{PD}. Above the blue line is the field-induced ferromagnetic phase. Below the red line is the labyrinth phase with a mixing of skyrmions and rectangular domains. The skyrmion crystal phase is found in a narrow region between these two lines, down to infinitesimal $D$.

\begin{figure}[h!]
\centering
\includegraphics[width=2 in]{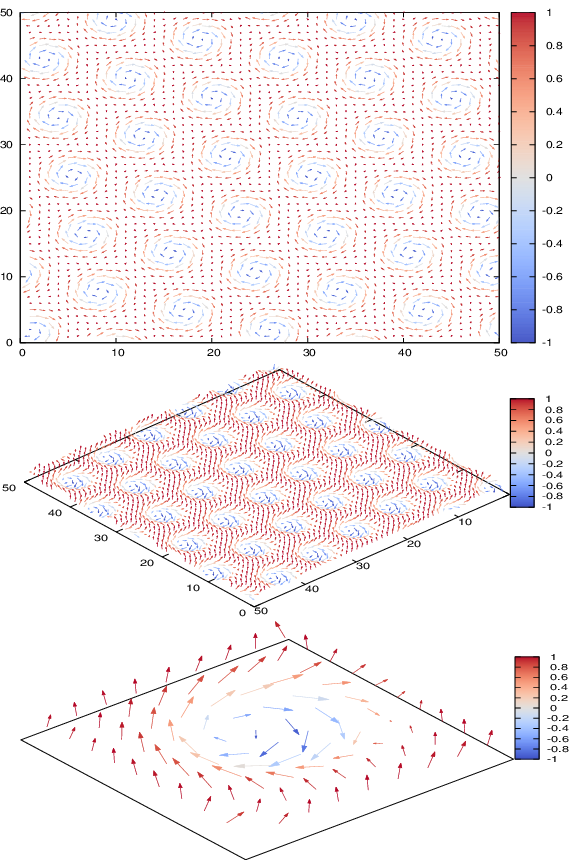}
\includegraphics[width=2 in]{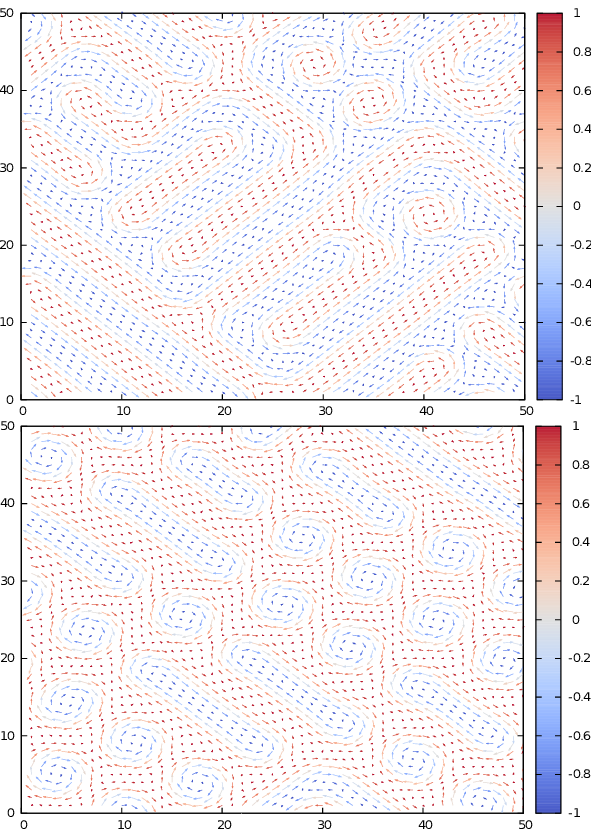}
\caption{Ground state for $D/J=1$ and $H/J=0.5$, a crystal of skyrmions is observed. Left: from above, skyrmion crystal viewed in the $xy$ plane,  a 3D view, zoom of the structure of a single vortex. The value of $S_z$ is indicated on the color scale. Right: Ground state for $D=1$ and $H=0$, a mixing of domains of long and
round islands (right top). Ground state for $D=1$ and $H= 0.25$, a mixing of
domains of long islands and vortices (right bottom). We call these structures the "labyrinth phase". \label{GS1}}
\end{figure}

Let us show an example of the skyrmion crystal observed at ($D=1,H=0.5)$ and
($D=0.5,H=0.15)$ (Fig. \ref{GS1} left). We see that the skyrmions form a crystal of triangular lattice.  The size of each skyrmion depends on the ratio $H/D$.

Below the red line of Fig. \ref{PD} is the labyrinth phase (Fig. \ref{GS1} right).

We wish to study the stability of the skyrmion crystal phase at finite $T$. To that end, we define an order parameter of the crystal as the projection of the actual spin configuration at the time $t$ at temperature $T$ on the GS configuration.  We should average this projection over a large number of Monte Carlo steps per spin. The order parameter  $M$ is

\begin{equation}\label{OP}
M(T)=\frac{1}{N^2(t_a-t_0)}\sum_i |\sum_{t=t_0}^{t_a} \mathbf S_i (T,t)\cdot \mathbf S_i^0(T=0)|
\end{equation}
where $\mathbf S_i (T,t)$ is the $i$-th spin at the time $t$, at temperature $T$, and $\mathbf S_i (T=0)$ is its state in the GS. The order parameter $M(T)$ is close to 1 at very low $T$ where each spin is only weakly deviated from its state in the GS. $M(T)$ is zero when every spin strongly fluctuates in the paramagnetic state.

\begin{figure}[h!]
\centering
\includegraphics[width=3 in]{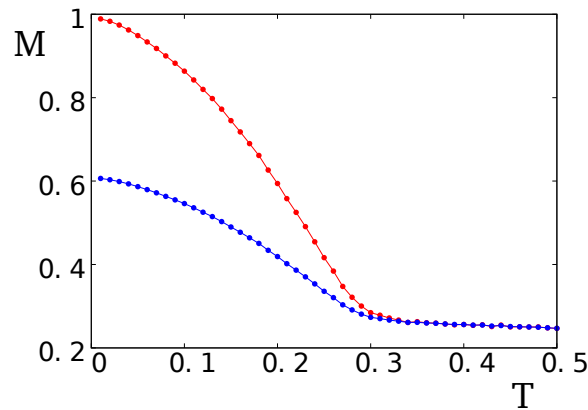}
\caption{Red circles: Order parameter defined in Eq. (\ref{OP}) versus $T$, for $H=0.5$ and
$N= 1800$. Blue crosses: the projection of the $S_z$ on $S_0^z$
of the ground state as
defined in Eq. (\ref{OP}) but for the $z$ components only. \label{MT}}
\end{figure}

We show in Fig. \ref{MT} the dependence of $M$ and $M_z$ on $T$ which indicate that for the skyrmion crystal remains ordered up to a finite $T$.  This stability at finite $T$ may be important for transport applications.

We have studied finite-size effects on the phase transition at $T_c$ and we have seen that from $N=800$, all curves coincide: there is no observable finite size effects for $N\geq 800$.

We have studied the relaxation time of skyrmions and found that it follows a stretched exponential law \cite{SahbiS2}.

The DM interaction has been shown above to generate a skyrmion crystal in a 2D lattice. But skyrmions  have been shown to exist in various kinds of lattices  \cite{Rosales, Buhrandt,Kwon1,Kwon2} and in crystal liquids \cite{Bogdanov2003,Leonov,Ackerman}. Experimental observations of skyrmion lattices have been realized in MnSi in 2009 \cite{Muhlbauer,Bauer} and in doped semiconductors in 2010 \cite{Yu1}. Also, the existence of skyrmion crystals have been found in thin films \cite{Ezawa,Yu2} and direct observation of the skyrmion Hall effect has been realized \cite{Jiang}. In addition, artificial skyrmion lattices have been devised for room temperatures \cite{Gilbert}.

It is noted that applications of skyrmions in spintronics have been largely discussed and  their advantages compared to early magnetic devices such as magnetic bubbles have been pointed out in a recent review by W. Kang {\it et al.}\cite{WangKang}. Among the most important applications of skyrmions, let us mention
skyrmion-based racetrack memory \cite{Parkin}, skyrmion-based logic gates \cite{Zhang,ZhouEzawa}, skyrmion-based
transistor \cite{XZhang2015b,JKim,YShiota} and skyrmion-based artificial synapse and neuron devices \cite{YHuang,SLi}.

Finally, we mention that we have found skyrmions confined at the interface
of a superlattice composed  alternately of a ferromagnetic film and an ferroelectric film \cite{Sharafullin,Sharafullin2}.  These results may have important applications.

\section{Conclusion}\label{Concl}
The aim of this review was to show a number of studied cases on the frustration and the surface effects in two dimensions and in thin films.  Interesting phenomena have been found to occur near the frontier of two competing phases of different ground-state orderings.  Without frustration, such frontiers do not exist. Exact solutions obtained for 2D Ising frustrated models show many striking features such as "partial disorder", namely a number of spins stay disordered in coexistence with ordered spins at equilibrium, "reentrance", namely a paramagnetic phase exists between two ordered phases in  a small region of temperature, and "disorder lines", namely lines on which the system looses one dimension to allow for a symmetry change from one side to the other.
Such beautiful phenomena can only be uncovered and understood by means of exact mathematical solutions. They allow us to understand qualitatively systems possessing similar microscopic ingredients but impossible to solve.

The surface effects have been studied by means of the Green's function method for frustrated non-collinear spin systems. Monte Carlo simulations have also used to elucidate many physical phenomena where analytical methods cannot be used. Surface spin-waves, surface magnetization and surface phase transition have been analyzed as functions of interactions, temperature and applied field.

We have also treated the question of surface criticality.  Results of our works show that critical exponents in thin films depend on the film thickness, their values lie between the values of 2D and 3D universality classes.

Recent results on skyrmions have also been reviewed in this paper. One of the striking findings is the discovery of a skyrmion crystal in a spin system with Dzyaloshinskii-Moriya interaction in competition with an exchange interaction, in a field.  This skyrmion crystal is shown to be stable at finite temperature.

To conclude, we would like to say that  investigations on the subjects discussed above continue intensively today. Note that there is an enormous number of investigations of other researchers on the above subjects and on other subjects concerning frustrated thin films. We  mentioned these works in our original papers, but to keep the paper length reasonable we did not present them here.

\section*{Acknowledgments}
The author wishes to thank his former doctorate students and collaborators for their close collaborations in the works presented in this review. In particular, he is grateful to Hector Giacomini, Patrick Azaria, Ngo Van Thanh,  Sahbi El Hog, Aur\'elien Bailly-Reyre and Ildus F. Sharafullin who have greatly contributed by their works to the understanding of frustrated thin films.


\begin{thebibliography}{999}
\bibitem{DiepFSS} H. T. Diep, Ed., {\it Frustrated Spin Systems}, 2nd ed., World Scientific, Singapore (2013).
\bibitem{DiepSP} H. T. Diep, {\it Statistical Physics- Fundamentals and Application to Condensed Matter}, World Scientific, Singapore (2015).
\bibitem{Zinn} J. Zinn-Justin, {\it Quantum Field Theory and Critical
Phenomena},
Oxford University Press, London, 4th edition (2002).

\bibitem{zangwill} A. Zangwill, {\it Physics at Surfaces},
Cambridge University Press, London (1988).

\bibitem{bland-heinrich} J.A.C. Bland and B. Heinrich (editors), {\it Ultrathin Magnetic Structures}, vol. I and
II,  Springer-Verlag, Berlin  (1994).

\bibitem{DiepTM} H. T. Diep, {\it Theory of Magnetism -Application to Surface Physics},  World Scientific, Singapore (2014).

\bibitem{Baibich} M. N. Baibich, J. M. Broto, A. Fert, F. Nguyen
Van Dau, F. Petroff, P. Etienne, G. Creuzet, A. Friederich and J.
Chazelas, Phys. Rev. Lett. {\bf 61}, 2472 (1988).

\bibitem{Grunberg} P. Grunberg, R. Schreiber, Y. Pang, M. B. Brodsky and
H. Sowers,
Phys. Rev. Lett. {\bf 57}, 2442 (1986); G. Binash, P. Grunberg, F.
Saurenbach and W. Zinn, Phys. Rev. B {\bf 39}, 4828 (1989).

\bibitem{Fert} A. Barth\'el\'emy et al, J. Mag.  Mag. Mater. {\bf
242-245}, 68 (2002).
\bibitem{Fert2013} A. Fert, V. Cros, and J. Sampaio,
Nat. Nanotechnol. {\bf 8}, 152 (2013).
\bibitem{Tou}  G. Toulouse,  Commun. Phys. {\bf 2},  115 (1977).
\bibitem{Villain1}  J. Villain,   J. Phys. C {\bf 10},  1717 (1977).
\bibitem {Wan} G. H. Wannier,  Phys. Rev. {\bf 79},  357 (1950);
 Phys. Rev. B {\bf 7}, 5017 (E) (1973).

\bibitem{Yos} A. Yoshimori,  J. Phys. Soc. Jpn. {\bf 14}, 807 (1959).

\bibitem{Vill} J. Villain,   Phys. Chem. Solids {\bf 11},  303 (1959).

\bibitem{Kapl} T. A. Kaplan,   Phys. Rev. {\bf 116},  888 (1959).

\bibitem{Dzyaloshinskii} I. E. Dzyaloshinskii, Thermodynamical Theory of 'Weak" Ferromagnetism in Antiferromagnetic Substances, Sov. Phys. JETP {\bf 5}, 1259 (1957).
\bibitem{Moriya} T. Moriya,  Anisotropic superexchange interaction and weak ferromagnetism, Phys. Rev. {\bf 120}, 91 (1960).

\bibitem{Berge}B. Berge, H. T. Diep, A. Ghazali, and P. Lallemand,
Phys. Rev. B  {\bf 34}, 3177 (1986).

\bibitem{Giacomini} H. T. Diep and H. Giacomini, Frustration - Exactly Solved Models, chapter 1, {\it Frustrated Spin Systems}, edited by H. T. Diep, pp. 1-58, World Scientific, Singapore (2013).


\bibitem{Bax} R. J. Baxter, {\it Exactly solved Models in Statistical
Mechanics\/}, Academic Press, New  York (1982).

\bibitem{Diep91a} H.T. Diep, M. Debauche and H. Giacomini,  Phys. Rev. B
{\bf 43}, 8759 (1991).
\bibitem{Ste} J. Stephenson,  J. Math. Phys. {\bf 11},  420 (1970);
 Can. J. Phys. {\bf 48},  2118 (1970);  Phys. Rev. B
{\bf 1}, 4405 (1970).
\bibitem{Rujan}P.  Rujan ,  J. Stat. Phys. {\bf 49}, 139 (1987) .
\bibitem{Mail}J. Maillard ,  Second conference on Statistical Mechanics,
California Davies (1986), unpublished.
\bibitem{Ka/Na} K. Kano and S. Naya,  Prog. Theor. Phys. {\bf
10},  158 (1953).

\bibitem{Aza87} P. Azaria, H. T. Diep and H. Giacomini,  Phys. Rev.
Lett. {\bf 59}, 1629 (1987).
\bibitem{Diep91b} M. Debauche, H.T. Diep, P. Azaria, and H. Giacomini,
 Phys. Rev. B {\bf 44},  2369 (1991).
\bibitem{Gaff} A. Gaff and J. Hijmann,  Physica A {\bf 80},  149 (1975).
\bibitem{Suzu} M. Suzuki
and M. Fisher,  J. Math. Phys. {\bf 12},  235 (1971).
\bibitem{Wu72} F. Y. Wu,  Solid Stat.
Comm. {\bf 10}, 115 (1972).

\bibitem{Sacco} J. E. Sacco and F. Y. Wu,  J. Phys. A {\bf
8}, 1780 (1975).

\bibitem{Diep92} M. Debauche and H. T. Diep,  Phys. Rev. B {\bf
46}, 8214 (1992); H. T. Diep, M. Debauche and H. Giacomini,
J. of Mag. and Mag. Mater. {\bf 104}, 184 (1992).

\bibitem{Horiguchi} T. Horiguchi,  Physica A {\bf 146}, 613 (1987).
\bibitem{Blan} D. Blankschtein, M. Ma and A. Berker,  Phys. Rev. B
{\bf 30},  1362 (1984).
\bibitem{Diep85b} H. T. Diep, P. Lallemand and O. Nagai,  J. Phys. C
{\bf 18}, 1067 (1985).
\bibitem{Blan85}D. Blankschtein, M. Ma , A. Nihat Berker,
G. S. Grest, and C. M. Soukoulis,  Phys. Rev. B {\bf 29},
5250 (1984).
\bibitem{Nagai} See the chapter by O. Nagai, T. Horiguchi
and S. Miyashita, in \cite{DiepFSS }.
\bibitem{Aza89b} P. Azaria, H. T. Diep and H. Giacomini,  Europhys. Lett.
{\bf 9},  755 (1989).

\bibitem{foster1} D.P. Foster, C. G\'erard and I. Puha,  J.
Phys. A: Math. Gen. {\bf 34}, 5183 (2001).

\bibitem{foster2} D.P. Foster and C. G\'erard,
Critical behavior of the fully frustrated q-state Potts piled-up-domino model,
Phys. Rev. B {\bf 70}, 014411 (2004).

\bibitem{Diep1979} Diep-The-Hung, J. C. S. Levy and O. Nagai, Effect of surface spin-waves and surface anisotropy in magnetic thin films at finite temperatures, Phys. Stat. Solidi (b) {\bf 93}, 351 (1979).

\bibitem{DiepTF91} H. T. Diep, Quantum effects in antiferromagnetic thin films,
Phys. Rev. B {\bf 43}, 8509 (1991).
\bibitem{Tyablikov} N. N. Bogolyubov and S. V. Tyablikov, Doklady Akad.
Nauk S.S.S.R. {\bf 126}, 53 (1959) [translation: Soviet
Phys.-Doklady  {\bf 4}, 604 (1959)].
\bibitem{Quartu1998} R. Quartu and H. T. Diep, Phase diagram of  body-centered tetragonal Helimagnets, J. Magn. Magn. Mater. {\bf 182}, 38 (1998).
\bibitem{Diep2015} H. T. Diep, Quantum Theory of Helimagnetic Thin Films, Phys. Rev. B {\bf 91}, 014436 (2015).

\bibitem{Mermin} N. D. Mermin and H. Wagner, Phys. Rev. Lett. {\bf 17}, 1133 (1966).

\bibitem{NgoSurface} V. Thanh Ngo and H. T. Diep, Effects of frustrated surface in Heisenberg thin films,
Phys. Rev. B {\bf 75}, 035412 (2007), Selected for the Vir. J. Nan. Sci. Tech. {\bf 15}, 126 (2007).

\bibitem{NgoSurface2} V. Thanh Ngo and H. T. Diep, Frustration effects in antiferrormagnetic face-centered cubic Heisenberg films,
J. Phys: Condens. Matter. {\bf 19}, 386202 (2007).

\bibitem{Ngo2007} V. T. Ngo and H. T. Diep, Phys. Rev. {\bf B75}, 035412  (2007).

\bibitem{Rocco} See, for example, R. Quartu and H.T. Diep,
Phys. Rev. B {\bf 55}, 2975 (1997).

\bibitem{santa2}C. Santamaria, R. Quartu and H. T. Diep,  J. Appl. Physics  {\bf 84},
1953 (1998).



\bibitem{Metropolis} N. Metropolis, A. W. Rosenbluth, M, N, Rosenbluth, and A. H. Teller, J. Chem. Phys. {\bf 21}, 1087 (1953).\\

\bibitem{Binder} K. Binder and D. W. Heermann,  {\it Monte Carlo Simulation in
Statistical Physics}, Springer-Verlag, Berlin, 1992.

\bibitem{Ferrenberg} A. M. Ferrenberg and R. H. Swendsen,
Phys. Rev. Let. 61(1988)2635; Phys. Rev. Let.  {\bf 63}, 1195(1989).
\bibitem{Ferrenberg2}
A. M. Ferrenberg and D. P. Landau, Phys. Rev. B {\bf} 44, 5081(1991).

\bibitem{Harada} I. Harada and K. Motizuki, J. Phys. Soc. Jpn 32(1972)927.

\bibitem{Diep1989a} H. T. Diep, Magnetic transitions in Helimagnets,
		Phys. Rev. B{\bf 39}, 397 (1989).
\bibitem{Diep1989b} H. T. Diep,  Low-temperature properties of quantum Heisenberg helimagnets
		Phys. Rev. B{\bf 40}, 741 (1989).

\bibitem{Diep2016} Sahbi El Hog and H. T. Diep, Helimagnetic Thin Films: Surface Reconstruction, Surface Spin-Waves, Magnetization,
 J. Magnetism and Magn. Mater.  {\bf 400}, 276-281 (2016).


\bibitem{DiepField2017} Sahbi El Hog and H. T. Diep, Partial Phase Transition and Quantum Effects in Helimagnetic Films under an Applied Field,
 J. Magnetism and Magnetic Materials {\bf 429}, 102 (2017).
\bibitem{DiepCritical} X. T. Pham Phu, V. Thanh Ngo and H. T. Diep, Critical Behavior of Magnetic Thin Films,
Surface Science  {\bf 603}, pp.109-116 (2009).


\bibitem{Fisher}  T. W. Capehart and M. E. Fisher,
 Phys. Rev.  B {\bf 13}, 5021 (1976).
\bibitem{Barber} M. N. Barber, Finite-Size Scaling, in: C. Domb and J. L. Lebowitz (Eds.), {\it Phase Transitions and Critical Phenomena}, Vol. 8, Academic Press, 1983, pp. 146-268.

\bibitem{Bunker}
A. Bunker, B. D. Gaulin, and C. Kallin, Phys. Rev. B {\bf  48},
15861 (1993).


\bibitem{DiepCrossover} X. T. Pham Phu, V. Thanh Ngo and H. T. Diep, Cross-Over from First to Second Order Transition in Frustrated Ising         Antiferromagnetic Films,
Phys. Rev. E {\bf 79}, 061106 (2009).

\bibitem{Skyrme} T. H. R. Skyrme, Proc. Roy. Soc. A {\bf 260} 127 (1961) ; A unified field theory of mesons and baryons, Nucl. Phys. {\bf 31}, 556 (1962).


\bibitem{Bogdanov2003} A. N. Bogdanov, U. K. R\"{o}${\ss}$ler and A. A. Shestakov, Skyrmions in liquid crystals, Phys. Rev. E {\bf 67}, 016602 (2003).

\bibitem{Leonov} A. O. Leonov, I. E. Dragumov, U. K. R\"{o}${\ss}$ler and A. N. Bogdanov, Theory of skyrmion states in liquid crystals, Phys. Rev. E {\bf 90}, 042502 (2014).

\bibitem{Ackerman} P. J. Ackerman, R. P. Trivedi, B. Senyuk, J. V. D. Lagemaat and I. I. Smalyukh, Phys. Rev. E {\bf 90}, 012505 (2014).

\bibitem{Ezawa} M. Ezawa, Giant Skyrmions Stabilized by Dipole-Dipole Interactions in Thin Ferromagnetic Films, Phys. Rev. Lett. {\bf 105}, 197202 (2010).

\bibitem{Yu2} X. Z. Yu, N. Kanazawa, Y. Onose, K. Kimoto, W. Z. Zhang,
S. Ishiwata, Y. Matsui, and Y. Tokura, Near room-temperature formation of a skyrmion crystal in thin-films of the helimagnet FeGe, Nature Mater. {\bf 10}, 106 (2011).

\bibitem{Yu1} X.Z. Yu, Y. Onose, N. Kanazawa, J. H. Park, J. H. Han, Y. Matsui, N. Nagaosa and Y. Tokura, Real-space observation of a two-dimensional skyrmion crystal, Nature {\bf 465} (7300), 901 (2010).
\bibitem{Muhlbauer} S. M\"{u}hlbauer, B. Binz, F. Jonietz, C. Pfleiderer, A. Rosch, A. Neubauer, R. Georgii and B. B\"{o}ni, Skyrmion Lattice in a Chiral Magnet, Science {\bf 323}, 915 (2009).

\bibitem{Bauer} A. Bauer and C. Pfleiderer, Magnetic phase diagram of MnSi inferred from magnetization and ac susceptibility, Phys. Rev. B {\bf 85}, 214418 (2012).

\bibitem{SahbiS} Sahbi El Hog, H. T. Diep and Henryk Puszkarski, Theory of Magnons in Spin Systems with Dzyaloshinskii-Moriya Interaction,
	           J. Phys. Condensed Matter {\bf 29}, 305001 (2017).

\bibitem{SahbiS2} Sahbi El Hog, Aur\'elien Bailly-Reyre and H. T. Diep, Stability and Phase Transition of Skyrmion Crystals Generated by Dzyaloshinskii-Moriya Interaction, J. Mag. Mag. Mater.  {\bf 455}, 32-38 (2018).

\bibitem{Rosales} H. D. Rosales, D. C. Cabra and P. Pujol, Three-sublattice Skyrmions crystal in the antiferromagnetic triangular lattice, Phys. Rev. B {\bf 92}, 214439 (2015).


\bibitem{Buhrandt} S. Buhrandt and L. Fritz, Skyrmion lattice phase in three-dimensional chiral magnets from Monte Carlo simulations, Phys. Rev. B {\bf 88}, 195137 (2013).

\bibitem{Kwon1}H. Y. Kwon, S. P. Kang, Y. Z. Wu and C. Won, Magnetic generated by Dzyaloshinskii-Moriya interaction, J. Appl. Phys. {\bf 113}, 133911 (2013).

\bibitem{Kwon2}H. Y. Kwon, K. M. Bu, Y. Z. Wu and C. Won, Effect of anisotropy and dipole interaction on long-range order magnetic structures generated by Dzyaloshinskii-Moriya interaction, J. Mag. Mag. Mater. {\bf 324}, 2171 (2012).


\bibitem{Jiang} Wanjun Jiang,	Xichao Zhang,	Guoqiang Yu,	Wei Zhang,	 Xiao Wang,	M. Benjamin Jungfleisch,	John E. Pearson,	Xuemei Cheng,	 Olle Heinonen,	 Kang L. Wang,	 Yan Zhou,	 Axel Hoffmann	and Suzanne G. E. te Velthuis, Direct observation of the skyrmion Hall effect, Nature     Physics {\bf 13}, 162-169 (2017).


\bibitem{Gilbert} D. A. Gilbert, B. B. Maranville, A. L. Balk, B. J. Kirby, P. Fischer, D. T. Piercen J. Unguris, J. A. Borchers and K. Liu, Realization of ground-state artificial skyrmion lattices at room temperature, Nature Comm.  {\bf 6}, Article number: 8462 (2015), published 8 Oct. 2015, DOI:10.1038/ncomms9462.

\bibitem{WangKang} Wang Kang, Yangqi Huang, Xichao Zhang, Yan Zhou, and
Weisheng Zhao, Skyrmion-Electronics: An
Overview and Outlook, Proceedings of the IEEE, August 2016,
DOI: 10.1109/JPROC.2016.2591578.

\bibitem{Parkin} S. S. P. Parkin, M. Hayashi, and
L. Thomas, Magnetic domain-wall
racetrack memory, Science {\bf 320},
no. 5873, pp. 190-194 (2008).


\bibitem{Zhang} Xichao Zhang, Motohiko Ezawa and Yan Zhou, Magnetic skyrmion logic gates: conversion, duplication and merging of skyrmions, Scientific Reports {\bf 5}, 9400 (2015).

\bibitem{ZhouEzawa} Y. Zhou and M. Ezawa, A reversible
conversion between a skyrmion and a
domain-wall pair in junction geometry,
Nat. Commun. {\bf 5}, Art. no. 4652 (2014).

\bibitem{XZhang2015b} X. Zhang, Y. Zhou, M. Ezawa,
G. P. Zhao, and W. Zhao, Magnetic
skyrmion transistor: skyrmion motion in a
voltage-gated nanotrack, Sci. Rep. {\bf 5}, Art. no. 11369 (2015).

\bibitem{JKim} J. Kim et al., Voltage controlled
propagating spin waves on a
perpendicularly magnetized nanowire,
 arXiv: 1401.6910 (2014).

\bibitem{YShiota} Y. Shiota et al., Quantitative evaluation of
voltage-induced magnetic anisotropy
change by magnetoresistance
measurement, Appl. Phys. Expr. {\bf 4},
no. 4, Art. no. 43005 (2011).

\bibitem{YHuang} Yangqi Huang, Wang Kang, Xichao Zhang, Yan Zhou and Weisheng Zhao,
Magnetic skyrmion-based synaptic devices, Nanotechnology {\bf 28}, 08LT02 (2017).

\bibitem{SLi} Sai Li, Wang Kang, Yangqi Huang, Xichao Zhang, Yan Zhou and Weisheng Zhao,
Magnetic skyrmion-based artificial neuron device,
Nanotechnology {\bf 28}, 31LT01 (2017).

\bibitem{Sharafullin} Ildus F. Sharafullin, Aidar G. Nugumanov, Alina R. Yuldasheva, Ainur R. Zharmukhametov, H. T. Diep, Modeling of magnetoelectric and surface properties in superlattices and nanofilms of multiferroics, to appear in J. Mag. Mag. Mater., https://doi.org/10.1016/j.jmmm.2018.11.116.

\bibitem{Sharafullin2}    Ildus F. Sharafullin, M. Kh. Kharrasov, H. T. Diep, Confined Skyrmions created by Dzyaloshinskii-Moriya Interaction at Interface of Magneto-Ferroelectric Superlattices, preprint.



\end{thebibliography}
\end{document}